\documentclass[%
preprint,
superscriptaddress,
amsmath,
amsymb,
aps]{revtex4-2}
\usepackage{natbib}
\usepackage{xcolor}
\usepackage{lingmacros}
\usepackage{tree-dvips}
\usepackage{xcolor}
\usepackage{graphicx}
\graphicspath{{../pdf/}{C:\Users\Shelby\OneDrive - US Navy-flankspeed\NRL\Lab\Altermagnets\Summaries\ORNL\Plots}}
\usepackage{epstopdf}
\usepackage{dcolumn}
\usepackage{bm}
\usepackage{gensymb}
\usepackage{upgreek}
\usepackage{hyperref}
\usepackage{tabularx}
\usepackage{float}
\usepackage{lineno}
\usepackage[absolute, overlay]{textpos}

\newcommand{\cameca}{CAMECA}

\modulolinenumbers[2]

\begin{document}

\begin{textblock*}{20cm}(3.5cm,26cm) 
   Distribution Statement A. Approved for public release: Distribution is unlimited
\end{textblock*}

\author{Shelby S. Fields}
\affiliation{Materials Science and Technology Division, U.S. Naval Research Laboratory, Washington, District of Columbia 20375, USA}

\author{Joseph C. Prestigiacomo}
\affiliation{Materials Science and Technology Division, U.S. Naval Research Laboratory, Washington, District of Columbia 20375, USA}

\author{Cory D. Cress}
\affiliation{Electronics Science and Technology Division, U.S. Naval Research Laboratory, Washington, District of Columbia 20375, USA}

\author{Nicholas G. Combs}
\affiliation{Materials Science and Technology Division, U.S. Naval Research Laboratory, Washington, District of Columbia 20375, USA}

\author{Olaf van 't Erve}
\affiliation{Materials Science and Technology Division, U.S. Naval Research Laboratory, Washington, District of Columbia 20375, USA}

\author{Patrick G. Callahan}
\affiliation{Materials Science and Technology Division, U.S. Naval Research Laboratory, Washington, District of Columbia 20375, USA}

\author{Keith E. Knipling}
\affiliation{Materials Science and Technology Division, U.S. Naval Research Laboratory, Washington, District of Columbia 20375, USA}

\author{Michelle E. Jamer}
\affiliation{Physics Department, United States Naval Academy, Annapolis, MD 21402, USA}

\author{Frank M. Abel}
\affiliation{Physics Department, United States Naval Academy, Annapolis, MD 21402, USA}

\author{Feng Ye}
\affiliation{Neutron Scattering Division, Oak Ridge National Laboratory, Oak Ridge, Tennessee 37831, USA}

\author{Arianna Minelli}
\affiliation{Neutron Scattering Division, Oak Ridge National Laboratory, Oak Ridge, Tennessee 37831, USA}

\author{Zachary J. Morgan}
\affiliation{Neutron Scattering Division, Oak Ridge National Laboratory, Oak Ridge, Tennessee 37831, USA}

\author{Haile Ambaye}
\affiliation{Neutron Scattering Division, Oak Ridge National Laboratory, Oak Ridge, Tennessee 37831, USA}

\author{Masaaki Matsuda}
\affiliation{Neutron Scattering Division, Oak Ridge National Laboratory, Oak Ridge, Tennessee 37831, USA}

\author{Avishek Maity}
\affiliation{Neutron Scattering Division, Oak Ridge National Laboratory, Oak Ridge, Tennessee 37831, USA}

\author{Valeria Lauter}
\affiliation{Neutron Scattering Division, Oak Ridge National Laboratory, Oak Ridge, Tennessee 37831, USA}

\author{Steven P. Bennett}
\affiliation{Materials Science and Technology Division, U.S. Naval Research Laboratory, Washington, District of Columbia 20375, USA}

\date{\today}
\begin{abstract}

Initially identified as a promising altermagnetic (AM) candidate, rutile RuO$_2$ has since become embroiled in controversy due to contradictory findings of modeling and measurements of the magnetic properties of bulk crystals and thin films. For example, despite observations of a bulk non-magnetic state using density functional theory, neutron scattering, and muon spin resonance measurements, patterned RuO$_2$ Hall bars and film heterostructures display magnetotransport signatures of magnetic ordering. Among the characteristics routinely cited as evidence for AM is the observation of exchange bias (EB) in an intimately contacted Fe-based ferromagnetic (FM) layer, which can arise due to interfacial coupling with a compensated antiferromagnet. Within this work, the origins of this EB coupling in Ru-capped RuO$_2$/Fe bilayers are investigated using polarized neutron diffraction, polarized neutron reflectometry, cross-sectional transmission electron microscopy, and super conducting quantum interference device measurements. These experiments reveal that the EB behavior is driven by the formation of an iron oxide interlayer containing Fe$_3$O$_4$ that undergoes a magnetic transition and pins interfacial moments within Fe at low temperature. These findings are confirmed by comparable measurements of Ni-based heterostructures, which do not display EB coupling, as well as magnetometry of additional Fe/Ru bilayers that display oxide-driven EB coupling despite the absence of the epitaxial RuO$_2$ layer. While these results do not directly refute the possibility of AM ordering in RuO$_2$ thin films, they reveal that EB, and related magnetotransport phenomena, cannot alone be considered evidence of this characteristic in the rutile structure due to interfacial chemical disorder.

\end{abstract}

\title{Non-Altermagnetic Origin of Exchange Bias Behaviors in Incoherent RuO$_2$/Fe Bilayer Heterostructures}
\pacs{}
\maketitle

\section*{Introduction}

	 Recent observations of a sublattice magnetic moment and above-ambient Ne\'el temperature,\cite{berlijnItinerantAntiferromagnetismRuO22017} as well as modeling reports of high (1.4 eV) band splitting,\cite{smejkalEmergingResearchLandscape2022} have generated tremendous interest in rutile RuO$_2$ as an altermagnetic (AM) candidate.\cite{liuInverseAltermagneticSpin2023,tschirnerSaturationAnomalousHall2023} Altermagnetism is a newly discovered class of magnetic ordering in which spin textures exist alongside fully compensated magnetic moments due to the breaking of both parity and time-reversal symmetry, resulting in electron spin polarization with no net magnetic moment.\cite{smejkalEmergingResearchLandscape2022,krempaskyAltermagneticLiftingKramers2024,smejkalCrystalTimereversalSymmetry2020,mazinAltermagnetismMnTeOrigin2023} These materials are poised to enable memory with switching in the THz frequencies that circumvents the stray field-induced scaling challenges hindering modern spintronic computing.\cite{dienyOpportunitiesChallengesSpintronics2020} Moreover, the preexisting development of RuO$_2$ for computing,\cite{fieldsWakeupFatigueMechanisms2021,vijayElectrodesPbZrxTi1xO3Ferroelectric1993} catalyst,\cite{stoerzingerOrientationDependentOxygenEvolution2017,chenMnDopedRuO2Nanocrystals2020} and energy storage\cite{asbaniUltrahighArealCapacitance2021} applications position it as a leading altermagnetic candidate for swift technological adoption. 
	 
Despite the promise, debate persists as to whether rutile RuO$_2$ intrinsically hosts altermagnetism. For example, while investigations of established antiferromagnets-turned-altermagnets, such as MnTe and CrSb, display definitive transport and band structure signatures of spin splitting,\cite{krempaskyAltermagneticLiftingKramers2024,reimersDirectObservationAltermagnetic2024} experimental observations of such characteristics for RuO$_2$ are less conclusive.\cite{baiObservationSpinSplitting2022,fengAnomalousHallEffect2022,tschirnerSaturationAnomalousHall2023,fengIncommensurateSpinDensity2024,wuFieldfreeSpinOrbit2025,hahnOffstoichiometricSurfaceReconstruction2025} In addition, numerous more recent first-principles calculations\cite{smolyanyukFragilityMagneticOrder2024a} and measurements of RuO$_2$ single crystals, including magnetotransport,\cite{pengUniversalScalingBehavior2025} Spin-Resolved/Angle-Resolved Photo Emission Spectroscopy (SARPES),\cite{liuAbsenceAltermagneticSpin2024} Muon Spin Resonance ($\mu$-SR), and neutron diffraction\cite{kesslerAbsenceMagneticOrder2024} have unambiguously reported a lack of sublattice moment in the rutile unit cell. As such, RuO$_2$ does not appear to intrinsically host compensated moments, and therefore cannot be an altermagnet in its ground state.

Several rationalizations have been proposed to explain the discrepancies between the unexpected magnetic behaviors observed in patterned thin films in contrast to bulk measurements and ground-state modeling. For example, density functional theory (DFT) modeling reports competing ferromagnetic (FM) and AM states in RuO$_2$ that is biaxially strained precisely to the in-plane lattice parameters of (110) TiO$_2$.\cite{wickramaratneEffectsAltermagneticOrder2025} Furthermore, first-principles calculations have shown that an AM state can also be stabilized through hole doping of RuO$_2$.\cite{smolyanyukFragilityMagneticOrder2024a} Significant strains as well as reductions in the number of electrons that may arise due to the presence of Ru vacancies or the intentional or unintentional inclusion of dopants can both readily occur in deposited films, rationalizing observations of AM properties in thin, patterned RuO$_2$ layers. For example, these predictions are supported by the observation of a spontaneous anomalous Hall effect in Ru$\it{_x}$Cr$_{1-}$$\it{_x}$O$_2$ (0.1 $\le$ $\it{x}$ $\le$ 0.3) grown on (110) TiO$_2$,\cite{wangEmergentZerofieldAnomalous2023} where Cr can serve as a hole dopant, although it should be noted that further modeling of this specific case has found\cite{smolyanyukOriginAnomalousHall2025} that added holes may remain bound to Cr ions within this alloy. DFT modeling has also identified that the RuO$_2$ (110) surface may host a magnetization,\cite{torunRoleMagnetismCatalysis2013} which penetrates $\sim$2 unit cells into the `bulk'.\cite{hoSymmetrybreakingInducedSurface2025} However, different surface terminations and orientations have not been investigated and the surfaces of deposited films are generally not atomically smooth, which may reduce the influence of this effect within a real-world device.

The theoretical investigation of the AM properties of RuO$_2$ using DFT modeling\cite{smejkalEmergingResearchLandscape2022,smejkalChiralMagnonsAltermagnetic2023,douAnisotropicSpinpolarizedConductivity2025} is frequently based upon the presupposed existence of antiferromagnetic (AFM) ordering.\cite{smolyanyukFragilityMagneticOrder2024a,wickramaratneEffectsAltermagneticOrder2025} Where experimental evidence supporting such properties in patterned RuO$_2$ thin films includes observations of nonlinear anomalous Hall behavior,\cite{tschirnerSaturationAnomalousHall2023,fengAnomalousHallEffect2022} magneto tunneling resistance,\cite{nohTunnelingMagnetoresistanceAltermagnetic2025a} spin splitting torque,\cite{zhangElectricalManipulationSpin2025} and THz emission\cite{liuInverseAltermagneticSpin2023}, a characteristic frequently cited as evidence for antiferromagnetism is the presence of exchange bias (EB, $\it{H}$$\rm{_{EB}}$) in RuO$_2$/FM heterostructures,\cite{fanConfinedMagnetizationSublatticeMatched2025,fengAnomalousHallEffect2022} which manifests as a shifted hysteresis loop arising due to exchange coupling at the interface.\cite{meiklejohnNewMagneticAnisotropy1957,jungblutExchangeBiasingMBEgrown1995,beaMechanismsExchangeBias2008} These behaviors have been observed in RuO$_2$ films intimately coupled with several different FM layers including SrRuO$_3$,\cite{fanConfinedMagnetizationSublatticeMatched2025} CoFe,\cite{fengAnomalousHallEffect2022} Co,\cite{wuFieldfreeSpinOrbit2025} Fe (this work), and NiFe.\cite{fengIncommensurateSpinDensity2024,abelProbingMagneticProperties2025} Among the aforementioned proposed explanations for antiferromagnetism in this rutile structure, no single mechanism accounts for the pervasiveness ($i.e.$ growth in several independent deposition chambers) and persistence of the effect at both coherent and incoherent RuO$_2$/FM interfaces, and in heterostructures with growth-imparted roughness, alluding to the possibility of several contributions. Most recently, this EB effect has been proposed to originate at the RuO$_2$/FM interface instead of the `bulk' of the RuO$_2$ film based upon Magneto-Raman measurements.\cite{abelProbingMagneticProperties2025} Potentially supporting the key role of the interface, interactions arising from low temperature transitions within interlayer oxides\cite{delaventaExchangeBiasInduced2012} or spin-glasses \cite{aliExchangeBiasUsing2007} have been shown to drive EB in other systems and may similarly contribute in RuO$_2$/FM structures, especially considering the low oxidation potential of atomic Ru compared to common transition metal magnets.\cite{shangEllinghamDiagramsBinary2024} In such cases, EB and related magnetotransport behavior\cite{bhatEnhancedSpinAccumulation2016} may be observed despite an absence of compensated moments or AM in RuO$_2$. Regardless, the presence of either an extrinsically stabilized surface magnetic state or a functional RuO$_2$/FM interface are both enabling for the application of RuO$_2$ in sensor and computing applications.\cite{linManipulatingExchangeBias2019} Moreover, understanding the mechanism governing this behavior may lead to its identification, disambiguation, and application in similar bilayer systems.

In this work, the origins of the EB effect in incoherent RuO$_2$/Fe/Ru heterostructures are investigated with a combination of superconducting quantum interference device (SQUID) magnetometry, polarized neutron reflectometry (PNR), and polarized neutron diffraction (PND) as well as atom probe tomography (APT), cross-sectional high-resolution transmission electron microscopy (HRTEM), and electron energy loss spectroscopy (EELS). While PND patterns of RuO$_2$ thin films contain no signatures of AFM ordering, PNR data and fitting reveal that the interface separating the RuO$_2$ and Fe layers hosts uncompensated moments that are canted within the plane of the film. The relative canting angles of the moments within the interfacial and FM layers, similar to the EB behavior widely observed using magnetometry, is sensitive to the field history of the heterostructure. APT, HRTEM/FFT analysis, and EELS of lamella prepared from this heterostructure find that the RuO$_2$/Fe interface consists of mainly iron oxide with a detectable content of magnetite (Fe$_3$O$_4$), which has been found to produce EB behaviors in other systems\cite{delaventaExchangeBiasInduced2012} due to a structural and electronic transition below 120~K,\cite{verweyElectronicConductionMagnetite1939,zhangElectronStatesMagnetism1991} as well as a Fe gradient into the RuO$_2$ layer. Further, EB behavior is shown through SQUID magnetometry to be absent in RuO$_2$/Ni heterostructures, which boast different oxidation potential considerations and stable compounds at the separating interface. Based upon these experiments, it is established that the EB observed in incoherently grown RuO$_2$/Fe-based FM heterostructures does not arise due to an AFM (or AM) ordering intrinsic to the sputtered heteroepitaxial RuO$_2$ layer. Instead, these behaviors are driven by the presence of a magnetically active interlayer that is stabilized by the high oxidation potential of Fe relative to Ru and influences the canting angle of ferromagnetic moments hosted by the interface.

\section*{Experimental Methods}
\subsection*{Material Synthesis}

Preparation of RuO$_2$ thin films, both alone on double-sided TiO$_2$ substrates (MTI) and as an embedded layer within TiO$_2$/RuO$_2$/FM/Ru heterostructures, was conducted using an AJA ATC Orion sputter deposition system. For this process, which has previously been shown to yield high-quality single-crystal films on (110) and (001)-oriented TiO$_2$,\cite{fieldsOrientationControlMosaicity2024} substrates were heated to 450~$^{\circ}$C in a 4~mTorr background pressure consisting of 7.5~sccm Ar and 7.5~sccm of O$_2$. After fifteen minutes of temperature equilibration, deposition occurred through the application of 1.48~W~cm$^{-2}$ of direct current (DC) power across a pure, 2-inch diameter Ru target (99.5\% purity, ACI Alloys) using a DCXS-750-4 Multiple Sputter Source supply. 

For thick samples prepared for neutron diffraction, 300~nm-thick films were grown on each polished side of five 10~mm $\times$ 10~mm (110)-oriented substrates, where extra TiO$_2$ pieces were used to separate sample surfaces from the platen. For heterostructures, 40~nm of RuO$_2$ was grown on 10~mm $\times$ 10~mm and 5~mm $\times$ 5~mm TiO$_2$ substrates with (110) and (001) orientations.  After, the substrates were allowed to cool to room temperature overnight in the deposition atmosphere. For heterostructure samples, two varieties of FM layer were subsequently grown without breaking vacuum at room temperature: Fe and Ni. Each material was deposited using a DC power of 7.9~W~cm$^{-2}$ applied across a pure target in the presence of 4~mTorr of Ar (15~sccm) using a MDX-1k power supply. A Ru capping layer was then deposited $in$ $situ$ at room temperature from the same target and power supply as RuO$_2$, but with 2.0~W~cm$^{-2}$ of power and in 3~mTorr of Ar (15~sccm). Additional (110) TiO$_2$/Fe/Ru bilayer control samples were prepared in a similar fashion, but without the deposition of the RuO$_2$ layer.

\subsection*{Structural Characterization}
Laboratory-based X-ray diffraction (XRD) characterization was conducted using a Bruker D8 Discover instrument equipped with a rotating anode Cu-K$\it{\alpha}$ source and an Eiger2 R 500K area detector. For thick samples prepared for neutron diffraction, high-resolution XRD (HRXRD) patterns were collected and asymmetric peaks were used to align the in-plane directions of each coupon before they were affixed together on the Al holder. For heterostructure samples, HRXRD and X-ray reflectivity (XRR) patterns were collected to confirm layer thicknesses and film quality. All HRXRD patterns were collected using a GE400 monochromator, 0.5~mm incident slit, and 0.4~mm diameter collimator, whereas all XRR measurements were made using a Goebel mirror and 0.5~mm collimator alone.

APT~\cite{kellyAtomProbeTomography2007,millerAtomProbeTomography2009, gaultAtomProbeMicroscopy2012, gaultAtomProbeTomography2021} was used to measure compositional variation with depth of the (110) TiO$_2$/RuO$_2$/Fe/Ru heterostructure. Specimens for APT were prepared through standard FIB lift-out and milling procedures~\cite{thompsonSituSitespecificSpecimen2007,millerReviewAtomProbe2007} using a Thermo Fisher Helios G3 DualBeam focused ion beam/scanning electron microscope (FIB/SEM) following sputtering of a thin protective Au layer on top of the sample. A \cameca\ 4000X Si local electrode atom probe (LEAP) employing a 355~nm ultraviolet pulsed laser, 60~pJ nominal laser pulse energy, a pulse repetition rate of 500~kHz, a 60~K specimen base temperature, and a detection rate of 0.005 ions per pulse (0.5\%) was used for data collection. Data reconstruction and analysis were performed using \cameca\ AP Suite version 6.3. Atoms detected within a central 30~mm diameter region of the position-sensitive detector, corresponding to approximately 20\% of the total detector area, were selected for analysis. Restricting the dataset to ions emitted near the apex of the APT specimen reduces the effects of specimen curvature, which can otherwise cause simultaneous sampling of multiple layers in a thin multilayer structure due to the curved projection of the tip surface. The analysis volume was reconstructed based on voltage evolution of the specimen radius, using an assumed evaporation field of 22~V~nm$^{-1}$. This value was chosen to produce an apparent Fe FM layer thickness of approximately 10~nm, matching the thickness measured by cross sectional HRTEM and X-ray and neutron reflectivity. Within this reconstructed volume, a 20~nm diameter cylindrical region of interest (ROI) was selected for further analysis. TEM foils were prepared with the same FIB-SEM instrument also using standard techniques.\cite{langfordPreparationTransmissionElectron2001} HRTEM micrographs were obtained on a JEOL F200 operating at 200 kV in TEM mode, whereas EELS maps were collected in scanning TEM (STEM) mode. EELS spectra were collected in dual EELS mode with a 0.15~eV resolution and a map size of 114~x~218 pixels, and a per-pixel dwell time of 0.02~s. The beam direction was aligned with the [001] direction in the TiO$_2$ layer. Analysis of the TEM micrographs was completed using Gatan DigitalMicrograph 3.61.4723.0, Fiji,\cite{schindelinFijiOpensourcePlatform2012} and in-house Python scripts.

\subsection*{Magnetic Property Measurement}
Temperature-resolved magnetic hysteresis loop ($\it{M}$~vs.~$\it{H}$) and magnetization versus time ($\it{M}$ vs. $\it{t}$) measurements were collected in a Quantum Design SQUID system. For EB measurements, the 5 mm $\times$ 5 mm heterostructures were first heated to 400 K (above the reported Ne\'el temperature\cite{berlijnItinerantAntiferromagnetismRuO22017,fengAnomalousHallEffect2022}) and a 1 T in-plane magnetic field was applied, followed by cooling down to 5~K, replicating procedures employed within other magnetotransport studies of RuO$_2$ heterostructures.\cite{nohTunnelingMagnetoresistanceAltermagnetic2025a,fengAnomalousHallEffect2022} Then, in-plane $\it{M}$~vs.~$\it{H}$ loops were measured on warming out to $\pm$~2~kOe starting at intervals of 10~K up to 100~K, and then 50~K up to room temperature. For $\it{M}$ vs. $\it{t}$ measurements, virgin samples were heated to 400~K in a 2~kOe in-plane magnetic field, and then magnetization was monitored for 16 hours. Volume magnetization adjustments were all completed using coupon area and magnetic layer thickness determined from X-ray reflectivity.

\subsection*{Neutron Scattering}
Several neutron-based techniques, including white-beam neutron diffraction, polarized neutron diffraction, and polarized neutron reflectometry were utilized to probe the magnetic behavior of the RuO$_2$ films and RuO$_2$/FM heterostructures at Oak Ridge National Laboratory. Room-temperature neutron diffraction measurements, performed at beamline BL-9 CORELLI at the Spallation Neutron Source (SNS), leveraged the white beam Laue technique employing a large coverage of position-sensitive detectors. To enhance signal, a set of five substrates with thick RuO$_2$ layers grown on both sides were stacked together on an Al support. The $\it{UB}$ orientation matrix was determined using the TiO$_2$ substrate by rotating the sample 360$\degree$ in 3$\degree$ increments around its rotational axis (aligned with the $c$-axis), which enabled the reconstruction of the reciprocal space maps. Finer steps of 0.2-0.4$\degree$ were utilized near the ($\bar{3}$$\bar{3}$0) and  (0$\bar{1}$0) reflections to study the wavelength-dependence of diffracted intensity. Polarized neutron diffraction measurements were conducted at the triple-axis instrument HB1 (PTAX) at the High-Flux Isotope Reactor using the same set of stacked thin film samples. For these measurements, a Heusler (111) monochromator and analyzer were utilized with a fixed incident neutron energy of 13.5 meV. Helmholtz coils were used to align the neutron spin along $x$ (along $Q$), $y$ (perpendicular to $Q$ in the scattering plane), and $z$ (vertical to the scattering plane) at the sample position. The horizontal collimator sequence was 48'-80'-sample-60'-open. The contamination from higher-order beams was eliminated using pyrolitic graphite filters. A cryofurnace was used to control the sample temperature for above-ambient (420~K) and low-temperature (35~K) measurements. Additional polarized neutron reflectometry measurements of a (110) TiO$_2$/RuO$_2$/Fe/Ru heterostructure were performed on the time-of-flight (ToF) Magnetism Reflectometer MAGREF\cite{lauterHighlightsMagnetismReflectometer2009} BL4A at the Spallation Neutron Source at Oak Ridge National Laboratory using neutrons with wavelengths ($\lambda$) between 0.26 and 0.86~nm and a high polarization of 98.5\%.\cite{tongSituPolarized3He2012,jiangNewGenerationHigh2017,syromyatnikovNewTypeWideangle2014} The experiments were conducted within a closed cycle refrigerator (Advanced Research Systems) in combination with a 1.15~T Bruker electromagnet.

\section*{Results and Discussion}

HRXRD 2$\theta$-$\omega$ patterns and rocking curve measurements made on (110) RuO$_2$ thin films are shown in Figure~1(a,b). HRXRD patterns collected on both the 300~nm (grown for neutron diffraction measurements) and 40~nm (grown within heterostructures) films, shown in Figure~1(a), display only peaks corresponding to the out-of-plane (110) direction, indicative of epitaxial templating from the TiO$_2$ substrate in both samples. However, the 300~nm-thick film boasts peak positions that are closer to bulk values, indicating that the thinner growth is expectedly more strained to the substrate and of higher crystalline quality [for comparison, the position of the bulk RuO$_2$ (110) peak is 28.14$\degree$ in 2$\it{\theta}$]. An additional comparison of quality is afforded through rocking curve measurements and full-width half maxima (FWHM) fitting, as displayed in Figure~1(b). Gaussian peak shapes fit to the rocking curve patterns from the 300~nm and 40~nm films display FWHM values of 0.572$\degree$ and 0.026$\degree$, respectively. The narrower FWHM of the 40~nm-thick sample is evidence of a greater crystallinity for this thickness, which further supports the superior quality of this film. It should be noted as well that mosaicity has been observed in (110) RuO$_2$ grown using an identical process to a thickness of 100~nm,\cite{fieldsOrientationControlMosaicity2024} which is likely also present in the 300~nm-thick sample. Equivalent data for 40~nm-thick (001) RuO$_2$, shown in supplemental information Figure~S1, similarly shows only a (002) out-of-plane film peak, which displays an $\omega$-width of 0.176$\degree$. The wider FWHM of the (002) peak is indicative of smaller crystalline domains in this growth than within the 40~nm-thick (110) film.

Further evaluation of the thicknesses, roughnesses, and densities of each layer present within each Fe and Ni heterostructure was completed using XRR, as detailed in supplemental Figure~S2. Less information is available in the fringes measured on the (001)-oriented samples than their (110) counterparts due to additional roughness in the RuO$_2$ layer. Overall, XRD and XRR measurements of each RuO$_2$ film indicate a high degree of epitaxial quality, where $\it{in}$ $\it{situ}$ growth has successfully produced heterostructures with discreet TiO$_2$/RuO$_2$/FM/Ru layers for examination of EB effects in RuO$_2$.

SQUID magnetometry measurements of Fe heterostructures were completed to confirm the presence of the EB behavior observed by other investigations.\cite{fengAnomalousHallEffect2022,fanConfinedMagnetizationSublatticeMatched2025,abelProbingMagneticProperties2025} An analysis of temperature-dependent $\it{M}$ vs. $\it{H}$ loops measured on TiO$_2$/RuO$_2$/Fe/Ru heterostructures with (110) and (001) TiO$_2$/RuO$_2$ orientations is shown in Figure~2(a-d). For both of the (110) and (001) orientations, shown in Figure~2(a) and (b), respectively, a clear shifting of the $\it{M}$ vs. $\it{H}$ is observed at lower temperatures. At 5~K, a shift of 1~kOe is observed for the (001)-oriented sample, which is larger than the 0.3~kOe shift observed for (110). By comparison, the loops measured on the (001)-oriented heterostructure are more rounded and display significantly larger coercive fields, which may be due to a larger degree of roughness present in the (001) RuO$_2$ surface.\cite{fieldsOrientationControlMosaicity2024,schragNeelOrangepeelCoupling2000,shahCompensationOrangepeelCoupling2014,zhaoEffectSurfaceRoughness2001} While this background effect may obscure potential differences related to the supposed Ne\'el vector difference between the two crystal orientations,\cite{fengAnomalousHallEffect2022} both clearly display EB shifting despite applied fields that are significantly lower than those reported to reorient Ne\'el vectors within the rutile structure.\cite{tschirnerSaturationAnomalousHall2023} The variation in positive and negative coercive fields (and their differences) with temperature are shown in Figure~2(c) and (d), respectively, for the (110) and (001) oriented heterostructures. In both cases, the coercive fields begin to shift at temperatures near 50~K, and in both cases the positive and negative coercive fields increase with decreasing temperature, with the latter changing more rapidly with field. Measurements of the (110) heterostructure following an in-plane rotation of 90$\degree$, shown in supplemental Figure~S3(a), also display this behavior. Further, a $\it{M}$ vs. $\it{H}$ measurement of the (110)-oriented heterostructure at 5~K following field cooling with a -1~T applied field from 400~K, shown in supplemental Figure~3(b), shows a shift in the opposite direction along the field axis, confirming that this shifting is programmable based upon the field cooling history. In total, these observations agree with prior investigations of magnetic coupling behavior of RuO$_2$/FM heterostructures, within which this shifting is cited as evidence of intrinsic compensated moments in rutile RuO$_2$,\cite{fengAnomalousHallEffect2022,fanConfinedMagnetizationSublatticeMatched2025} or an extrinsic coupling occurring at the interface.\cite{abelProbingMagneticProperties2025}

To investigate whether the observed EB effect occurs due to defect-driven magnetic ordering present within RuO$_2$ thin films,\cite{smolyanyukFragilityMagneticOrder2024a} white-beam and polarized neutron diffraction data were collected on a stacked set of five RuO$_2$/TiO$_2$ (110 substrate)/RuO$_2$ samples, as shown in Figure~3(a-e). It should be noted that the current body of literature on neutron diffraction investigations of RuO$_2$ contains contradiction, where initial reports of a forbidden nuclear peak that persisted to above room temperature\cite{berlijnItinerantAntiferromagnetismRuO22017} have since been disputed by additional measurements that attribute this intensity to a multiple scattering event.\cite{kesslerAbsenceMagneticOrder2024} Moreover, extrinsic contributions to the magnetic ordering in RuO$_2$ including defects such as Ru vacancies\cite{smolyanyukFragilityMagneticOrder2024a} or dopants (intentional or unintentional),\cite{wangEmergentZerofieldAnomalous2023} both of which may be present in sputtered thin films, have been proposed to explain discrepancies between bulk measurements and transport data. Accordingly, neutron diffraction of RuO$_2$ thin films on non-magnetic TiO$_2$ substrates affords the opportunity to investigate whether the EB effect in RuO$_2$/FM heterostructures occurs due to the presence of a defect-stabilized sublattice moment in sputtered RuO$_2$ films, as well as the potential contributions of multiple scattering to forbidden diffraction peaks in neutron diffraction patterns of this rutile structure.

A room-temperature neutron diffraction pattern of the (110) stacked sample along the (001) zone-axis is shown in Figure~3(a). Strong diffraction spots from the large volume of TiO$_2$ substrate are present as well as weaker and more diffuse spots at reciprocal vectors corresponding to the stacked RuO$_2$ films in their vicinity. A high-resolution map of the (150) peak, an allowed nuclear reflection in both structures, is shown in Figure~3(b). Despite a low intensity, the (150) RuO$_2$ peak is clearly visible at a slightly larger scattering vector than the corresponding TiO$_2$ peak, which indicates that the stacked sample contains enough RuO$_2$ film volume for observation using neutron diffraction. Further, the pattern shown in Figure~3(a) contains forbidden ($\it{h}$00) and (0$\it{k}$0) ($\it{h}$ or $\it{k}$ = odd) peaks, which have been previously noted as evidence of magnetic ordering in RuO$_2$.\cite{berlijnItinerantAntiferromagnetismRuO22017} A high-resolution map of the (0$\bar{3}$0) peak, shown in Figure~3(c) reveals that this forbidden peak corresponds to the TiO$_2$ substrate owing to its $\it{d}$-spacing, shape, and intensity, and further that no intensity is present in the vicinity from the RuO$_2$ film stack. Given that TiO$_2$ is non-magnetic, it is evident that multiple scattering is responsible for the presence of the forbidden peaks in this pattern. A further confirmation of this is obtained through wavelength-dependent measurements, as described in the supplemental information and shown in Figure~S4(a,b). While a lack of RuO$_2$ intensity in this forbidden region could be interpreted as a lack of sublattice moment in this structure, the comparatively small film volume, likely lower quality of the RuO$_2$ films compared to the TiO$_2$ substrates, and potentially small sublattice moment on Ru make it difficult to conclusively determine, using this non-polarized room-temperature measurement alone, if sputtered RuO$_2$ thin films are non-magnetic. Accordingly, above-ambient and low-temperature polarized neutron diffraction measurements were further undertaken using a triple-axis spectrometer to investigate potential magnetic contributions to allowed, nuclear low-$\it{Q}$ RuO$_2$ peaks.

Polarization-dependent $\omega$ rocking scans of the (110) RuO$_2$ stacked sample collected above (420~K) and below (35~K) the EB transition temperature using a polarized triple-axis neutron spectrometer are shown in Figures~3(d) and 3(e), respectively. Due to the low intensity from the available RuO$_2$ peaks and their proximity to the stronger TiO$_2$ substrate peaks, it was necessary to perform $\omega$ rocking curve scans through the RuO$_2$ peaks for integrated intensity comparisons between the different channels. An example 2$\theta$-$\omega$ scan collected around the (220) TiO$_2$/RuO$_2$ peaks at 35~K that displays the difference in intensity and position between the two reflections is shown in supplemental Figure~S5.  The use of a polarized neutron beam to probe RuO$_2$ $\omega$ peaks enables the investigation of any present compensated magnetic ordering, where a preferred spin orientation would be expected due to anisotropic strains, crystalline anisotropy, and film relaxation. Accordingly, if such ordering is present and its magnetic structure factor allows intensity at the $\omega$ peak position, then any spin orientation projected along the $\it{x}$, $\it{y}$, or $\it{z}$ direction (within the sample reference frame) should produce more scattered intensity when a measurement is made using the corresponding polarization channels. Therefore, comparison between the integrated intensities in different channels with spin flip (SF) and non-spin flip (NSF) neutrons provides a means to investigate whether or not RuO$_2$ hosts compensated magnetic ordering.  At both 420~K and 35~K, integrated intensities of (110) peaks collected using all three SF polarization channels, $P_{\rm{x}}$, $P_{\rm{y}}$, and $P_{\rm{z}}$ are identical at each temperature, with no deviations outside of the error bars, which supports that no detectable magnetic moment is present within the stacked sample. Using the fit error associated with the available counting statistics in these measurements, a lack of sublattice moment detection places an upper limit of 0.01~$\mu$$_\text{B}$ per Ru atom in sputtered RuO$_2$ thin films, which is above a more sensitive muon spin resonance analysis limit of 0.00075~$\mu$$_\text{B}$ per Ru atom obtained by Ke$\ss$ler $\it{et}$ $\it{al.}$\cite{kesslerAbsenceMagneticOrder2024} while reporting a lack of detectable sublattice moment. In the context of the EB behavior observed in the TiO$_2$/RuO$_2$/Fe/Ru heterostructures, these neutron diffraction measurements support that magnetic ordering due to the presence of defects from film deposition is not a contributor to this effect, which is more likely driven by an interfacial interaction with the FM layer.

To explore the depth-resolved magnetic behavior of the RuO$_2$/Fe interface, sequential PNR measurements of the (110) TiO$_2$/RuO$_2$/Fe/Ru heterostructure were carried out. In a ToF PNR experiment, a highly collimated polarized neutron beam with a wavelength band ($\Delta$$\lambda$) impinges on the film at a grazing angle $\theta$, interacting with atomic nuclei and the spins of unpaired electrons. The reflected intensities, $R$$^+$ and $R$$^-$, are measured as a function of momentum transfer, $Q$ = 4$\pi$sin($\theta$)/$\lambda$, with the neutron spin parallel (+) or antiparallel (-) to the applied field. Shown in Figure~4(a-c) are PNR profiles collected at 5~K on this sample following field cooling from 400~K in a 1~T in-plane magnetic field, similar to the procedure that was employed to observe EB behavior using SQUID magnetometry. In these plots, the dark-color data and fits ($R$$^+$) correspond to the measurement with the neutron spin parallel, whereas the light-color data and fits ($R$$^-$) correspond to measurement with the neutron spin antiparallel to the direction of the applied field. Splitting between the $R$$^+$ and $R$$^-$ profiles occurs due to scattering of the incident neutrons off of the aligned, uncompensated moments within the sample stack. Accordingly, these reflectivity profiles and their splitting contain information about the depth profiles of chemical and magnetization vector distributions throughout the layers as well as their angle relative to the direction of the applied field and the neutron polarization.\cite{lauter-pasyukNeutronGrazingIncidence2007} Thus, measured experimental reflectivities $R$$^+$ and $R$$^-$ contain non-spin-flip ($R$$^{++}$ and $R$$^{--}$) and spin-flip ($R$$^{+-}$ and $R$$^{-+}$) components. $R$$^{++}$ and $R$$^{--}$ are determined by the component of the magnetization vector parallel to the direction of the neutron polarization and $R$$^{+-}$ and $R$$^{-+}$ are determined by the perpendicular component of the magnetization vector. This is accounted for within the fit so that $R$$^+$ = $R$$^{++}$ + $R$$^{+-}$ and $R$$^{--}$ = $R$$^{--}$ + $R$$^{-+}$.\cite{toperverg3DPolarizationAnalysis2001,blundellSpinorientationDependenceNeutron1995} Through fitting, scattering length density profiles are obtained for both nuclear (NSLD) and magnetic (MSLD) components of the films as well as MSLD canting angle profiles, as shown in Figure~4(d).\cite{cressDomainStateExchange2023} The initial measurement, shown in Figure~4(a), was conducted at 5~K in a 1~T field after the field cooling procedure, followed by a measurement at remanence [27~Oe, Figure 4(b)], and then a measurement at remanence (27~Oe) after application of a switching field of -1~T [Figure 4(c)]. To obtain the most reliable fits, identical NSLD and MSLD profiles were used for each dataset (with multiplication of the MSLD by -1 for the measurement made following switching) and only the canting angle of each magnetic layer was allowed to vary.

The MSLD and NSLD profiles obtained by fitting each PNR measurement, shown in Figure~4(d) in grey and black, respectively, reveal the details of the nuclear and magnetic structure at the interface between the RuO$_2$ and Fe layers. Where the Fe layer displays an expected MSLD of  5.4~$\times$~10$^{-6}$~\AA $^{-2}$, it is flanked by bottom [labeled `Int. 1' in Figure~4(d) and (e)] and top [labeled `Int. 2' in Figure~4(d)] interfaces that host smaller MSLDs of 9.5~$\times$~10$^{-7}$ and 1.5~$\times$~10$^{-6}$~\AA $^{-2}$, respectively. In addition, the NSLD of Int. 1 is lower than that of either of the intentional RuO$_2$ or Fe layers, which indicates that this layer has formed through chemical intermixing, in particular supporting a reduction of the RuO$_2$ surface. The in-plane canting angles of each magnetic layer, detailed on the bottom of Figure~4(d), all vary with measurement coordinate, where the angles of the moments present within Int. 1 display the largest absolute angular magnitudes of between 135 and 105$^\circ$. Figure~4(e) displays the evolution of the canting angle of the in-plane moments within the Fe and Int. 1 layers with measurement coordinate. Following cooling in a 1~T field and measurement in the same, the moments within Int. 1 display a canting angle of 105$^\circ$. After reduction of the applied field to remanence, this angle increases 135$^\circ$, followed by a magnitude reduction to -123$^\circ$ at remanence (27~Oe) after application of a -1~T field. These angular changes in Int. 1 occur in contrast to Fe, which is conversely well aligned with the direction of the applied 1~T (or -1~T) field. At saturation the moments within this layer cant by an angle of 3$\degree$, followed by an increase to 13$\degree$ as the applied field is reduced to remanence. After application of a -1~T field, measurement at remanence reveals a monotonic canting angle switch of -13$\degree$. The difference between the canting angles of these moments and those present within Int. 1 do not switch monotonically, which is a manifestation the EB that here is shown to be sensitive to the direction of the cooling field. Moreover, this variation in the canting angles of the moments hosted by Fe and Int. 1 represent the only changes required to obtain fits to each dataset, the quality of which is additionally confirmed by spin asymmetry fitting shown in supplemental Figure~S6(a). Further, intrinsic differences between the remanent measurements are evident upon inspection of their spin asymmetry difference, shown in supplemental Figure~S6(b), in which oscillations at large $Q$ values are indicative of a magnetic difference between the two fits present within a thin layer. While PNR cannot detect the presence of antiferromagnetically or speromagnetically\cite{coeyCharacterisationMagneticProperties1973,coeyNewSpinStructure1973} coupled moments, it is clear, owing to the sensitivity of the canted moments within Fe and Int. 1 to the field history of the heterostructure, that magnetic coupling at the RuO$_2$/Fe interface is responsible for pinning moments within the Fe film, which has been shown to mechanistically drive exchange biases in other systems.\cite{fitzsimmonsPinnedMagnetizationAntiferromagnet2007,ohldagCorrelationExchangeBias2003}

Based upon the presence of field history-sensitive canted moments at the RuO$_2$/Fe interface, it can be concluded that there exists a short-range magnetic interaction between these two layers that drives the observed exchange bias. Given that the presence of the canted moments within Int. 1 coincide with an unexpected decrease in NLSD, it is also evident that mass transfer plays a role in this interaction. To interrogate this RuO$_2$/Fe interface in the (110) TiO$_2$/RuO$_2$/Fe/Ru heterostructures, APT was conducted on a cylindrical lamella prepared using a focused ion beam (FIB).  Shown in Figure~5(a) and 5(b) are an APT reconstruction and composition depth profile, respectively, of a 20~nm-diameter region of interest (ROI) from the (110) TiO$_2$/RuO$_2$/Fe/Ru heterostructure. The interfaces are qualitatively delineated as reference for comparison with Figure~4(d) and (e). Discrete atom maps for different pure Ru and Fe, shown in supplemental information Figure~S7(a) and 7(b), respectively, provide additional information on the oxidation state of these interfaces. Comparison between the composition profiles in Figure~5(b) and the atomic reconstructions in Figure~S7(a) and S7(b) reveals that the RuO$_2$/Fe interface contains both a discrete reduced Ru layer, confirming observations made using PNR, and an oxidized Fe gradient that decays into the RuO$_2$ film, which are both indicative of mass transport. Further, oxidation is also observed at the interface separating Fe and Ru, which likely arises due to a combination of roughness in the polycrystalline films as well as imperfect oxidation protection from the capping layer, the latter of which is further supported by the decaying profile of adventitious carbon into the Ru surface. These observations are additionally supported by high-resolution tunneling electron microscopy (HRTEM) characterization of the film and interface layers, which is shown in Figure~5(c). All layers, including Int. 1 and Int. 2, display thicknesses that are in agreement values obtained from PNR fitting, independently verifying the PNR fit quality. Both interfacial layers further display unique contrast in TEM compared with Ru, Fe, or RuO$_2$, which additionally supports that interfacial oxides have formed between these intentional layers.

A further examination of oxide layer formation and mass transport between the Fe and RuO$_2$ layers is provided by integration of Fe-${L}$$_{\rm3}$ and O-$K$ EELS intensity through the interface, as shown in Figure~5(d), which is correlated with local atomic abundance. Fe-${L}$$_{\rm3}$ intensities are at a maximum within the Fe layer and decay into both Int. 2 and Int. 1, with the profile of the latter displaying an intensity tail that penetrates into the RuO$_2$, which is qualitatively matched by the Fe APT profile shown in Figure~5(b). O-$K$ intensities display local maxima within Int. 1 and Int. 2, as well as a relatively constant intensity in the RuO$_2$ layer, further confirming the presence of oxide interfaces observed through APT. A substantial O-$K$ intensity is also measured the Fe layer, which likely arises due to interface roughness from Int. 1 and Int. 2. While unambiguous determination of the oxide phases present within the interfaces is non-trivial due to the small physical volume, atomic columns are visible in localized regions of Int. 1 and Int. 2 within the HRTEM micrograph, which are analyzable using fast Fourier transform (FFT). An integrated FFT pattern of the combined RuO$_2$, Fe, and Int. 1 layers, shown in Figure~5(e), displays strong peaks that correspond to the heteroepitaxial RuO$_2$ layer, as well as diffuse intensity spots corresponding to $\alpha$-Fe. Peaks at locations consistent with forbidden reflections in RuO$_2$ [(100) and (200)] are also visible, which may arise due to biaxial strain within this layer or due to point defects from diffusion or reduction. Additional intensity is also observed at a reciprocal spacing of 0.39~$\rm{\AA}$$^{-1}$, which corresponds to the highest intensity (311) reflection of magnetite ($F$$d$$\bar{3}$$m$ Fe$_3$O$_4$). While the limited intensity and microscope capabilities prevent observation of structural transformations in this layer with temperature, the qualitative presence of diffraction spots at this spacing supports the formation of Fe$_3$O$_4$ at the interface, which critically has been shown to drive exchange biases in Fe-based thin films\cite{delaventaExchangeBiasInduced2012} and nanomaterials\cite{ongExchangeBiasFe2009,ongRoleFrozenSpins2011} through a Verwey transition. It is important to note that characteristic Verwey behavior involves both structural and metal-insulator transitions in magnetite, which in the context of EB leads to an easy axis rotation,\cite{delaventaExchangeBiasInduced2012} whereas here it is only the ambient-temperature presence of the $F$$d$$\bar{3}$$m$ Fe$_3$O$_4$ structure at the RuO$_2$/Fe interface that is experimentally observable, albeit with limited intensity. Accordingly, it is difficult to unambiguously determine the exact interaction mechanism between Int. 1 and the Fe layer, especially considering the potential contributions to the magnetic behavior that may accompany a spin-glass transition from the Fe gradient in RuO$_2$.\cite{aliExchangeBiasUsing2007} Regardless, HRTEM characterization of the interface separating the Fe and RuO$_2$ layers confirms the presence of a reaction interlayer that contains magnetically active Fe$_3$O$_4$, which has in other cases been shown to drive exchange biases through a low-temperature structural and electronic transition.

It is evident, based upon PNR measurements, APT profiles, and cross-sectional HRTEM/FFT analysis and EELS of the interface separating the RuO$_2$ and Fe layers, that chemical intermixing results in the formation of an interlayer that contains Fe$_3$O$_4$ within TiO$_2$/RuO$_2$/Fe/Ru heterostructures. However, confirming that an iron oxide layer drives this EB requires investigation of the heterostructure absent this interface. While preparation of an identical heterostructure without chemical intermixing is non-trivial, the use of Ni as the ferromagnet within this stack, which boasts a lower oxidation potential than Fe\cite{charetteThermodynamicPropertiesOxides1968} and a different collection of stable oxide compounds,\cite{shangEllinghamDiagramsBinary2024} enables the examination of this effect in the presence of a different interface. SQUID magnetometry measurements of Ni-based heterostructures with (110) and (001)-oriented TiO$_2$/RuO$_2$ made following field cooling in 1~T from 400~K are shown in Figure~6(a) and 6(b), respectively. Despite a field cooling procedure identical to the Fe-based heterostructures, loops measured down to 5~K on the Ni-based stacks do not show any significant shifting along the field axis. As plotted in Figure~6(c) and 6(d) for the (110) and (001)-oriented heterostructures, respectively, coercive fields for both samples increase with decreasing temperature. However, these increases are matched and $\it{M}$ vs. $\it{H}$ loop centers remain stagnant near $H$ = 0~kOe. The same is true for identical measurements performed on the (110)-oriented sample rotated 90$\degree$ in-plane, as shown in supplemental Figure~S8. A lack of EB in the Ni-based heterostructures confirms that the mechanism is not intrinsically related to RuO$_2$ or its surface, and is instead driven by a chemical interaction specifically with the Fe layer. This observation is further supported by identical $\it{M}$ vs. $\it{H}$ and related coercive field  measurements, shown in Figure~6(e) and 6(f), respectively, of a Fe/Ru bilayer stack deposited directly on (110) TiO$_2$ without the RuO$_2$ layer. $\it{M}$ vs. $\it{H}$ loops of this bilayer sample begin to shift to lower fields around 50~K, with centers falling to -0.15~kOe by 5~K. While the onset temperature of this shift is almost identical to that observed within the equivalent RuO$_2$/Fe heterostructure, the magnitude of the shift is less significant. As the stable TiO$_2$ substrate\cite{shangEllinghamDiagramsBinary2024} is unlikely to significantly oxidize the polycrystalline Fe film, oxidation through the protective Ru layer, which was observed through APT and HRTEM/EELS of the complete (110)-oriented heterostructure (Int. 2 in Figure~4 and Figure~5), is likely driving the EB observed in this sample. As such, the lower EB magnitude in the Fe/Ru bilayer compared to the RuO$_2$/Fe heterostructure supports that the unique oxidation conditions at the interface between the heteroepitaxial RuO$_2$ layer and the Fe produces a stronger magnetic interaction, possibly through enhanced Fe$_3$O$_4$ crystallinity or more homogenous chemical ordering. Regardless, interfacial reactions between Fe (and Fe-based FMs) and oxide layers\cite{fanExchangeBiasInterface2013,delaventaExchangeBiasInduced2012} have been observed to drive exchange-biases using magnetometry, supporting this mechanism for RuO$_2$. In particular, comparison of magnetic behavior of Al$_2$O$_3$/Fe$_{0.8}$Ni$_{0.2}$/V$_2$O$_3$ and Al$_2$O$_3$/Ni/V$_2$O$_3$ bilayers by De La Venta $et$ $al.$ revealed an exchange bias only in the presence of Fe$_{0.8}$Ni$_{0.2}$, which was attributed to a Verwey transition from the formation of a Fe$_3$O$_4$ interfacial layer.\cite{delaventaExchangeBiasInduced2012} While this study measured both positive shifts in magnetic moment and concomitant shifting of coercive field due to easy-axis rotation in Fe$_3$O$_4$ , a lack of EB in a Ni-based bilayer is similar to the result obtained here, validating the use of different ferromagnetic layers to assess the role of chemical intermixing in this effect in RuO$_2$/Fe heterostructures.

Additional measurements of the change in saturated moment with time ($\it{M}$ vs. $\it{t}$) at 400~K of both heterostructures as well as the bilayer sample, detailed in Figure~7, provide information on the formation of chemically intermixed interfaces within these film stacks. As all samples are held at 400~K, saturated moments decrease, with the RuO$_2$/Fe-based heterostructure experiencing a more rapid, nonlinear decrease compared to the linear reduction of the Ni-based heterostructure and Fe/Ru bilayer. It is reasonable to assume that the origin of each decrease is oxidation, given that the oxides and suboxides of Fe and Ni display smaller magnetizations than their parent elemental materials and that the curie points of both ferromagnets are well above 400~K. As such, normalized magnetizations can be taken as a proxy for oxidation,\cite{rebodosEffectsOxidationMagnetization2010,johnston-peckSynthesisStructuralMagnetic2009} where a `complete' reaction corresponds to some value that is arbitrarily lower than the normalized maximum. In this context, the linear negative slopes of the $\it{M}$ vs. $\it{t}$ of the Ni-based heterostructure and Fe/Ru bilayer correspond to oxidation processes that obey zeroth-order kinetics, which are not rate limited by the oxygen concentration at the interface of the RuO$_2$ or TiO$_2$ and do not proceed through intermediate compounds. Such behavior would be expected for the RuO$_2$/Ni heterostructure if the formed Ni oxide was less oxidized than RuO$_2$, which is the case for the singularly thermodynamically stable compound NiO. Similarly, the Fe oxidation within the Fe/Ru bilayer is limited by the content of oxygen available at the Fe/Ru interface given the relative stability of the TiO$_2$ substrate. In contrast, the nonlinear $\it{M}$ vs. $\it{t}$ behavior of the Fe-based heterostructure supports higher-order oxidation kinetics at the RuO$_2$ interface. For example, a plot of the inverse of the normalized saturated moment versus time reveals a nearly linear slope, which is indicative that the oxidation within the Fe layer proceeds through an intermediate suboxide, is rate-limited by the concentration of oxygen at the RuO$_2$ surface, or some combination thereof. In both cases, the RuO$_2$ may act as an oxygen reservoir, from which iron oxidation at the interface may proceed through suboxide intermediate compounds.

Based upon magnetic, chemical, structural, and kinetic experiments examining the interface separating RuO$_2$/Fe and RuO$_2$/Ni, it can be concluded that EB effects observed in RuO$_2$/Fe-based FM heterostructures occur due to a chemical interaction with Fe and formation of a reaction layer. Explored directly using sequential PNR measurements, this EB proceeds via a magnetic coupling between compensated and uncompensated moments at the RuO$_2$/Fe interface, the latter of which display an observable canting angle that possesses a projection with the Fe layer that is sensitive to the field history. While cross-sectional microscopy of this oxidized interface reveals the presence of Fe$_3$O$_4$, the EB coupling begins to become observable around 50~K, which is cooler than would be expected for a thermodynamic Verwey transition (115 - 120~K). However, the influences of sub-stoichiometry,\cite{aragonInfluenceNonstoichiometryVerwey1985} inclusions\cite{brabersImpurityEffectsVerwey1998} (in this case, likely Ru), and scaling\cite{leeSizeDependenceMetal2015,mitraVerweyTransitionUltrasmallSized2014,goyaStaticDynamicMagnetic2003} have all been individually shown to reduce this transition to temperatures below 100~K, and would likely be present in concert within a thin, thermally grown metal/oxide interface. Regardless, a lack of EB behavior in Ni-based heterostructures and the presence of EB in a Fe/Ru bilayer lacking the RuO$_2$/Fe interface confirm that intrinsic compensated magnetic ordering in RuO$_2$ is not responsible for this behavior observed in incoherent RuO$_2$/Fe-based FM heterostructures.

\section*{Conclusions}

Rutile RuO$_2$ has been the subject of intense investigation and development toward emergent altermagnetic applications following recent reports of a sublattice moment that persists above ambient temperatures as well as predictions of large band splitting. While recent calculations and bulk measurements have found that the structure is non-magnetic, experiments exploring the magnetotransport properties of patterned thin films stand in contrast to these results. Among the characteristics that are commonly referenced as evidence supporting altermagnetism in the rutile structure in patterned thin films is the observation of EB coupling, and associated hysteresis loop shifting following field cooling, of a FM layer in intimate contact with the RuO$_2$ surface. Here, through SQUID magnetometry in concert with PND, PNR, APT, and HRTEM/EELS, the nature of this exchange bias coupling has been explored. This characterization reveals that sputtered RuO$_2$ thin films do not host compensated sublattice moments, and that chemical disorder at the RuO$_2$/Fe interface drives the EB effect through interfacial moment pinning. Supporting this conclusion, this EB was not observable when Ni was used as the FM layer, likely due to its different oxidation potential, and was observable within TiO$_2$/Fe/Ru bilayers (prepared without the RuO$_2$ layer) due to oxidation through the protective Ru barrier. Based upon these observations, it is concluded that EB in RuO$_2$/Fe-based FM heterostructures, and associated magnetotransport phenomenon, should not be considered, alone, as evidence of AFM or AM within RuO$_2$. This conclusion does not refute the possibility of altermagnetism within the rutile structure, however care must be taken to isolate AM or AFM effects from those of layers that may form due to intimate contact with RuO$_2$ at modest temperatures.

\section*{Acknowledgement}

The authors would like to thank Dr. Darshana Wickramaratne and Dr. Marc Currie at the U.S. Naval Research Laboratory for productive discussions. This work was supported by the Office of Naval Research 6.1 Base Funding at the U.S. Naval Research Laboratory in Washington, D.C. Measurements at Oak Ridge National Laboratory (ORNL) were supported by the U.S. Department of Energy (DOE), Office of Science, Office of Basic Energy Sciences, and the U.S. DOE, Office of Science User Facility operated by ORNL. The beam time was allocated to CORELLI and PTAX on proposal number IPTS-32719 and to MAGREF on proposal IPTS-32732. Notice: This manuscript has been authored by UT-Battelle, LLC under Contract No. DE-AC05-00OR22725 with the U.S. Department of Energy. The United States Government retains and the publisher, by accepting the article for publication, acknowledges that the United States Government retains a non-exclusive, paid-up, irrevocable, world-wide license to publish or reproduce the published form of this manuscript, or allow others to do so, for United States Government purposes. The Department of Energy will provide public access to these results of federally sponsored research in accordance with the DOE Public Access Plan.\cite{USDepartmentEnergy2023} Research at the United States Naval Academy was supported by the Kinnear Fellowship and Office of Naval Research under Contract No. N0001423WX02132 and continued ONR support.

\bibliographystyle{unsrt}

\bibliography{References_list_tex_3}
\label{References}
\newpage

 \begin{figure}[h!]
 \centering
  \includegraphics[width=0.4\textwidth]{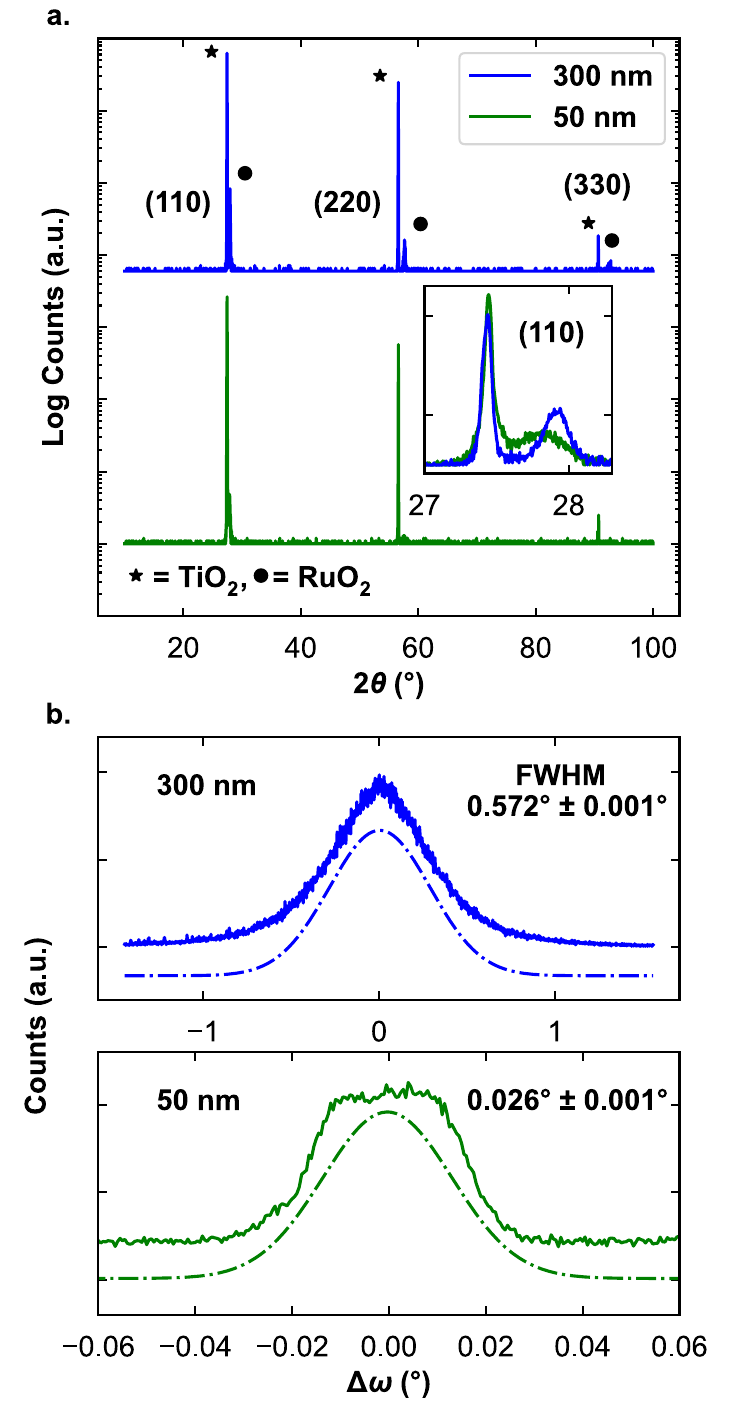}
 \caption{(a) High resolution XRD patterns and (b) (110) rocking curve peaks collected on 300 nm (blue) and 40 nm (green) thick epitaxial RuO$_2$ films grown on (110)-oriented TiO$_2$ substrates. In panel (a), the reflections are labeled and the originating materials are indicated using star (TiO$_2$) and closed circle (RuO$_2$) points, whereas in panel (b), the FWHM of each peak is labeled. Note the different $\Delta$$\omega$ axis limits in the subpanels of (b).}
 \end{figure}

 \begin{figure}[!h]
 \centering
  \includegraphics[width=0.7\textwidth]{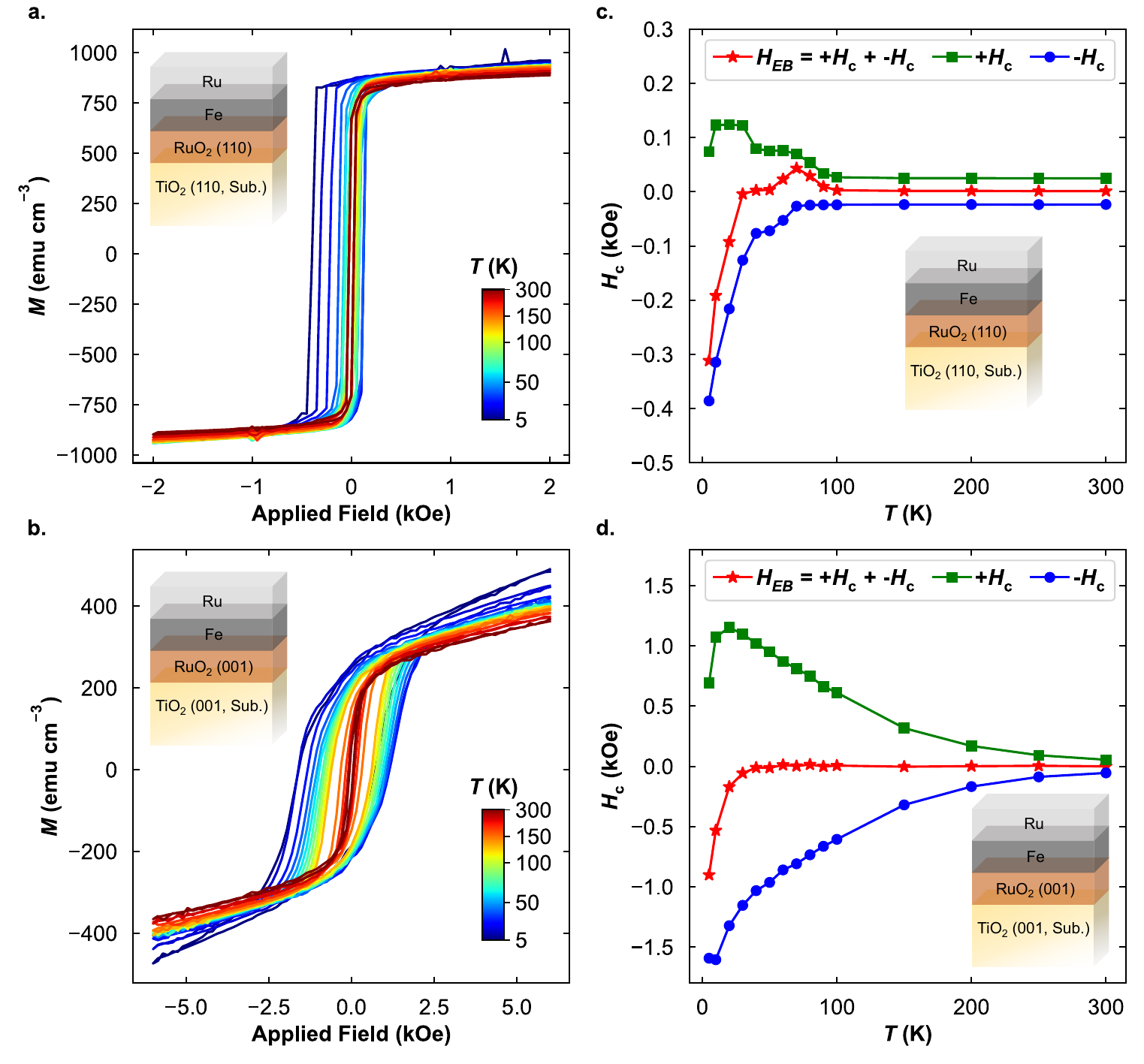}
 \caption{Temperature-dependent in-plane hysteresis loops measured on TiO$_2$/RuO$_2$/Fe/Ru heterostructures with (a) (110) and (b) (001)-oriented RuO$_2$ layers on heating following field cooling from 400 K in a 1 T applied field. Warmer colors correspond to higher $\it{T}$, whereas cooler colors correspond to lower, as detailed in the inset color bars. Positive (green, squares), negative (blue, circles), coercive fields as well as bias fields (red, stars) extracted from hysteresis loops measured on (c) (110) and (d) (001) TiO$_2$/RuO$_2$/Fe/Ru heterostructures. Inset within each panel is an annotated diagram of the measured film stack including TiO$_2$/RuO$_2$ orientation.}
 \end{figure}

 \begin{figure}[h!]
 \centering
  \includegraphics[width=0.8\textwidth]{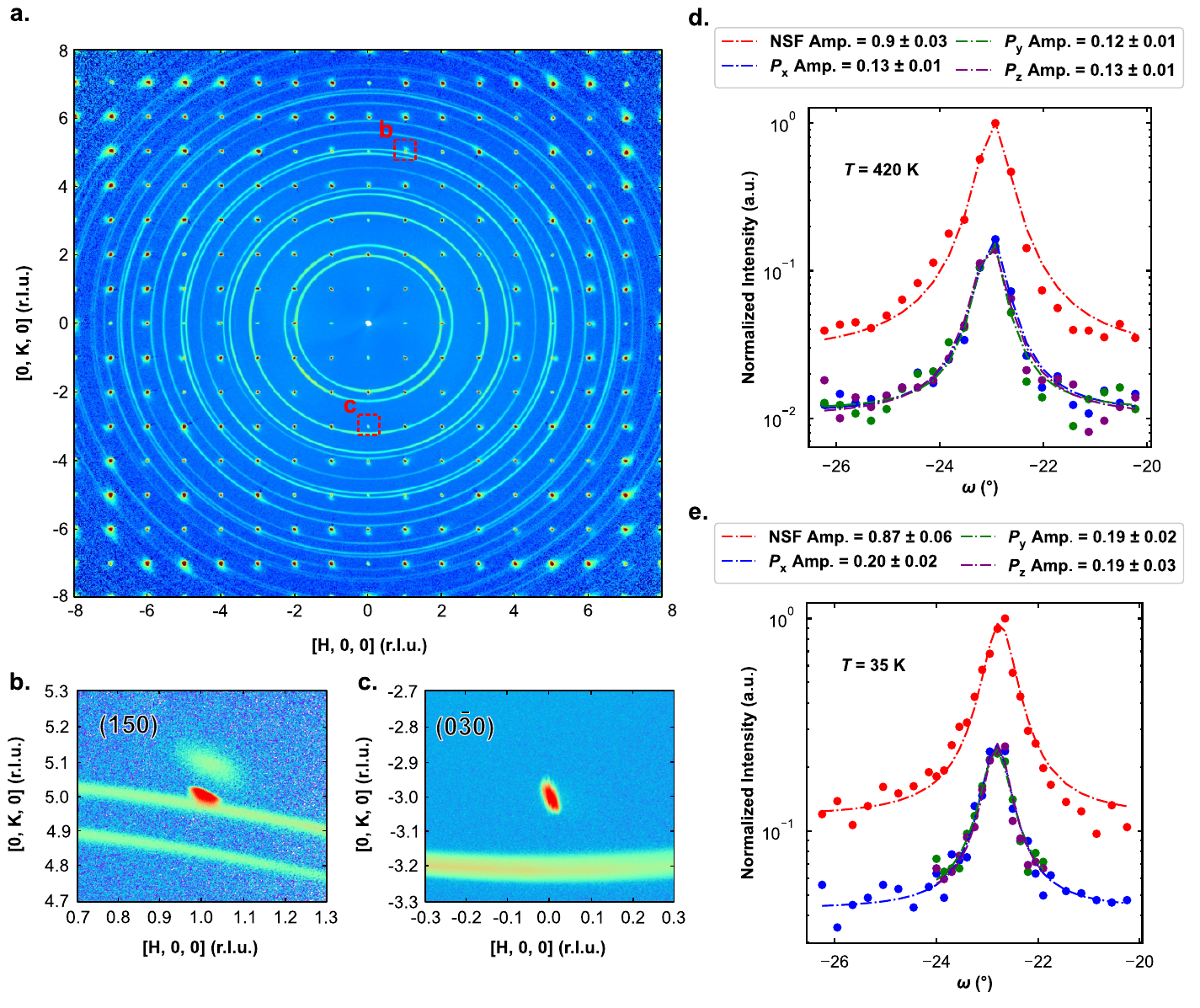}
 \caption{(a) White-beam neutron diffraction pattern measured at room temperature on a stack of (110)-oriented RuO$_2$ thin films, where the utilized $\it{UB}$ orientation matrix corresponds to TiO$_2$. Intensity rings arise due to the Al support used to place the sample in the instrument. Within this pattern, the (0$\bar{3}$0) and (150) peaks are identified with red dotted boxes and annotated with their corresponding panels, (b) and (c), which below display high-resolution reconstruction maps of these regions. Panels (d) and (e) display polarized neutron rocking scans of the (110) RuO$_2$ peaks collected at 420~K and 35~K, respectively. In these plots, points correspond to measured data whereas color-coded broken lines correspond to Lorentzian fits, and red (NSF, $P_{\rm{x}}$), blue (SF, $\it{P}$$_x$), green (SF, $\it{P}$$_y$), and purple (SF, $\it{P}$$_z$) indicate the orientation of the neutron spin at the sample position, as detailed in each legend. The legends also contain information on integrated intensities for the different polarization channels from the fitted Lorentzian amplitudes of each peak.}
 \end{figure}

 \begin{figure*}[!ht]
 \centering
  \includegraphics[width=1\textwidth]{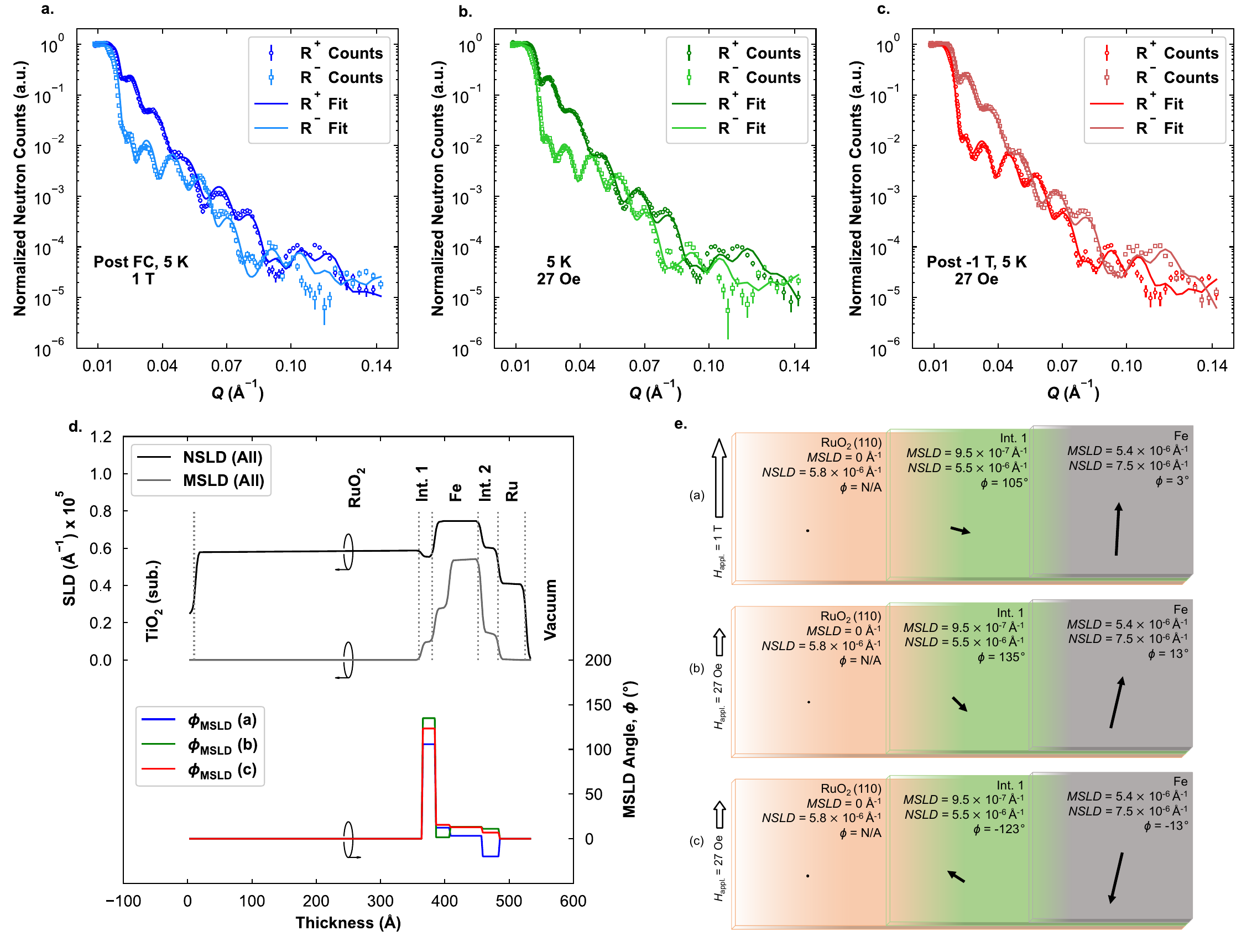}
 \caption{Sequential polarized neutron reflectometry measurements (open points) and fits (lines) of the (110) TiO$_2$/RuO$_2$/Fe/Ru heterostructure made following field cooling from 400 K in a 1 T field (a) at 5 K in a 1 T field, (b) at 5 K in a 27 Oe field, and (c) at 5 K in a 27 Oe field following application of a -1 T field. In these panels, dark and light colors correspond to measurements and fits with the polarizer on and off, respectively. (d) NSLD (black) and MSLD (gray) profiles (top data, left $y$-axis) obtained from the fits shown in panels (a), (b), and (c), along with the fit in-plane rotations ($\phi$) of the moments of each layer (bottom data, right $y$-axis), shown in blue, green, and red, respectively, with the red data multiplied by -1 to accommodate the negative MSLD. A diagram of the heterostructure during each measurement is provided in panel (e), where the corresponding data, layer, MSLD, NSLD, and in-plane canting angle (filled arrow, $\phi$) are annotated along with the measurement field (open arrow, left).}
 \end{figure*}

 \begin{figure*}[!ht]
 \centering
  \includegraphics[width=0.8\textwidth]{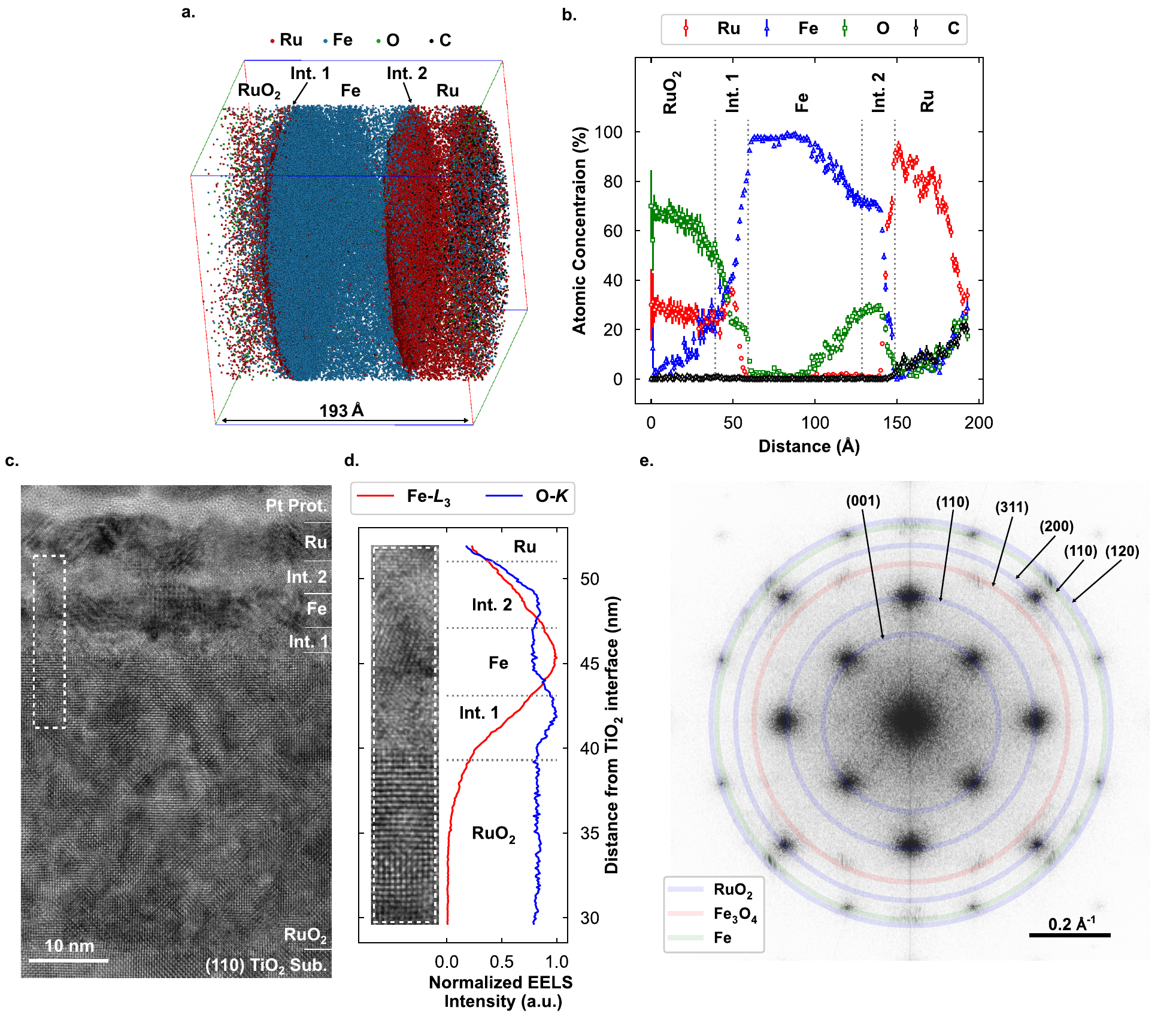}
 \caption{(a) Atomic reconstruction of APT data from a cylindrical lamella prepared using the (110) TiO$_2$/RuO$_2$/Fe/Ru surface. Layers are identified as well as interfaces corresponding to Figure 4(d) and (e). Red, blue, green, and black circles correspond to Ru, Fe, O, and C atoms, respectively, as indicated above the reconstruction. (b) Atomic concentration depth profiles integrated from the reconstruction in panel (a), where layer identities are listed at the top of the panel. Red circles, blue triangles, green squares, and black diamonds correspond to Ru, Fe, O, and C, respectively. (c) HRTEM micrograph of FIB-prepared lamella. The layers are detailed on the right side of the micrograph. (d) Fe-$\it{L}$$_{\rm3}$ (red) and O-$\it{K}$ (blue) EELS intensity integrated as a function of depth within the white dotted rectangle in panel (c). The rough positions of the regions detailed in panel (c) are listed to the right, while to the left is a zoom-in of the HRTEM micrograph from panel (c) showing the integrated region. (e) FFT of area of a HRTEM micrograph containing the Fe layer, Int. 1, and RuO$_2$ layer where dark color corresponds to intensity. Indexes for rutile $P$4$_2$/$mnm$ RuO$_2$, $F$$d$$\bar{3}$$m$ Fe$_3$O$_4$, and $I$$m$$\bar{3}$$m$ Fe are shown as blue, red, and green rings, respectively, with reflection indexes annotated using arrows.}
 \end{figure*}

 \begin{figure}[!ht]
 \centering
  \includegraphics[width=0.7\textwidth]{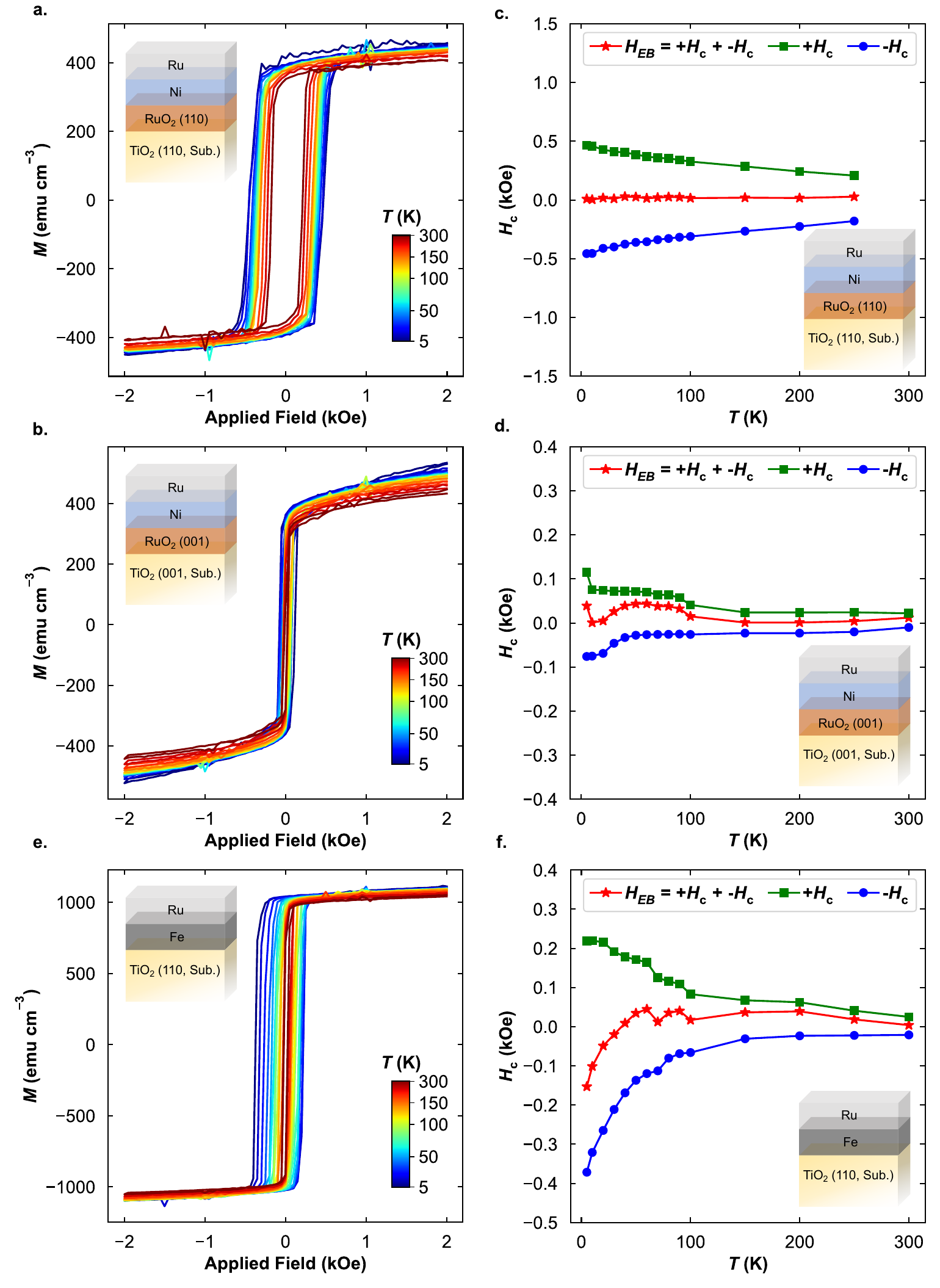}
 \caption{Temperature-dependent in-plane hysteresis loops measured on control samples comprising TiO$_2$/RuO$_2$/Ni/Ru heterostructures with (a) (110) and (b) (001)-oriented RuO$_2$ layers on warming following field cooling from 400 K in a 1 T applied field. Warmer colors correspond to higher $\it{T}$, whereas cooler colors correspond to lower, as detailed in the inset color bars. Positive (green, squares), negative (blue, circles), coercive fields as well as bias fields (red, stars) extracted from hysteresis loops measured on (c) (110) and (d) (001) TiO$_2$/RuO$_2$/Ni/Ru heterostructures. (e) Hysteresis loop and associated (f) coercive and exchange bias fields measured on a (110) TiO$_2$/Fe/Ru bilayer on warming following an identical field cooling procedure to the samples in panels (a-d). Inset within each panel is an annotated diagram of the measured film stack including TiO$_2$/RuO$_2$ orientation.}
 \end{figure}

 \begin{figure}[!ht]
 \centering
  \includegraphics[width=0.4\textwidth]{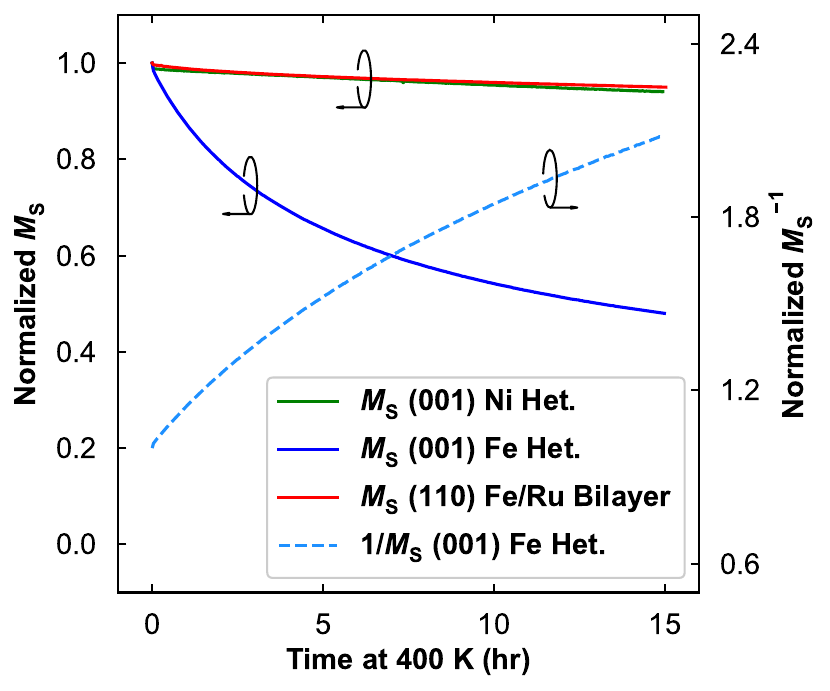}
 \caption{$\it{M}$ vs. $\it{t}$ measurements showing evolution of $M$$_{\rm{s}}$ of (001) TiO$_2$/RuO$_2$/Ni/Ru (green), (001) TiO$_2$/RuO$_2$/Fe/Ru (blue) heterostructure, and (110) TiO$_2$/Fe/Ru bilayer coupons with time at 400 K and 2 kOe, where moments have been normalized for ease of comparison. $\it{M}$$^{-1}$ vs. $\it{t}$ for the (001) TiO$_2$/RuO$_2$/Fe/Ru data is plotted as a light blue dashed line (corresponding to the right $\it{y}$-axis).}
 \end{figure}

\clearpage

\setcounter{figure}{0}
\renewcommand \thesection{S\arabic{section}}
\renewcommand\thetable{S\arabic{table}}
\renewcommand \thefigure{S\arabic{figure}}

\begin{textblock*}{20cm}(7.5cm,3cm) 
   Supplemental Information for:
\end{textblock*}

\begin{textblock*}{20cm}(5cm,3.5cm) 
   Non-Altermagnetic Origin of Exchange Bias Behaviors in
\end{textblock*}

\begin{textblock*}{20cm}(6cm,4cm) 
   Incoherent RuO$_2$/Fe Bilayer Heterostructures
\end{textblock*}

\vspace*{3cm}

\begin{figure}[!ht]
 \centering
  \includegraphics[width=0.4\textwidth]{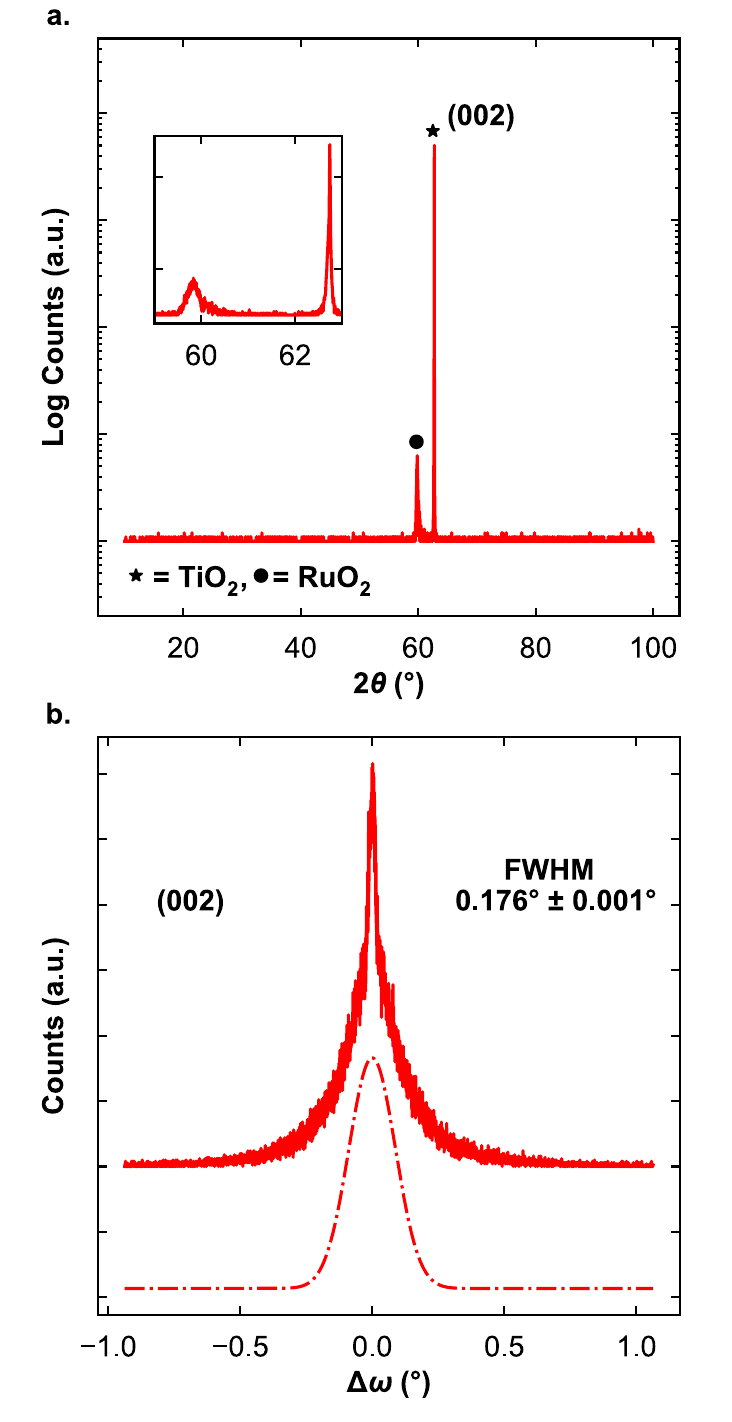}
 \caption{(a) High resolution XRD pattern and (b) (002) rocking curve peak collected on a 40~nm thick RuO$_2$ film grown on a (001)-oriented TiO$_2$ substrate. In panel (a), the reflections are labeled and the originating materials are indicated using star (TiO$_2$) and closed circle (RuO$_2$) points, whereas in panel (b), the FWHM of the rocking curve peak is labeled.}
 \end{figure}
 
 \begin{figure}[!ht]
 \centering
  \includegraphics[width=0.8\textwidth]{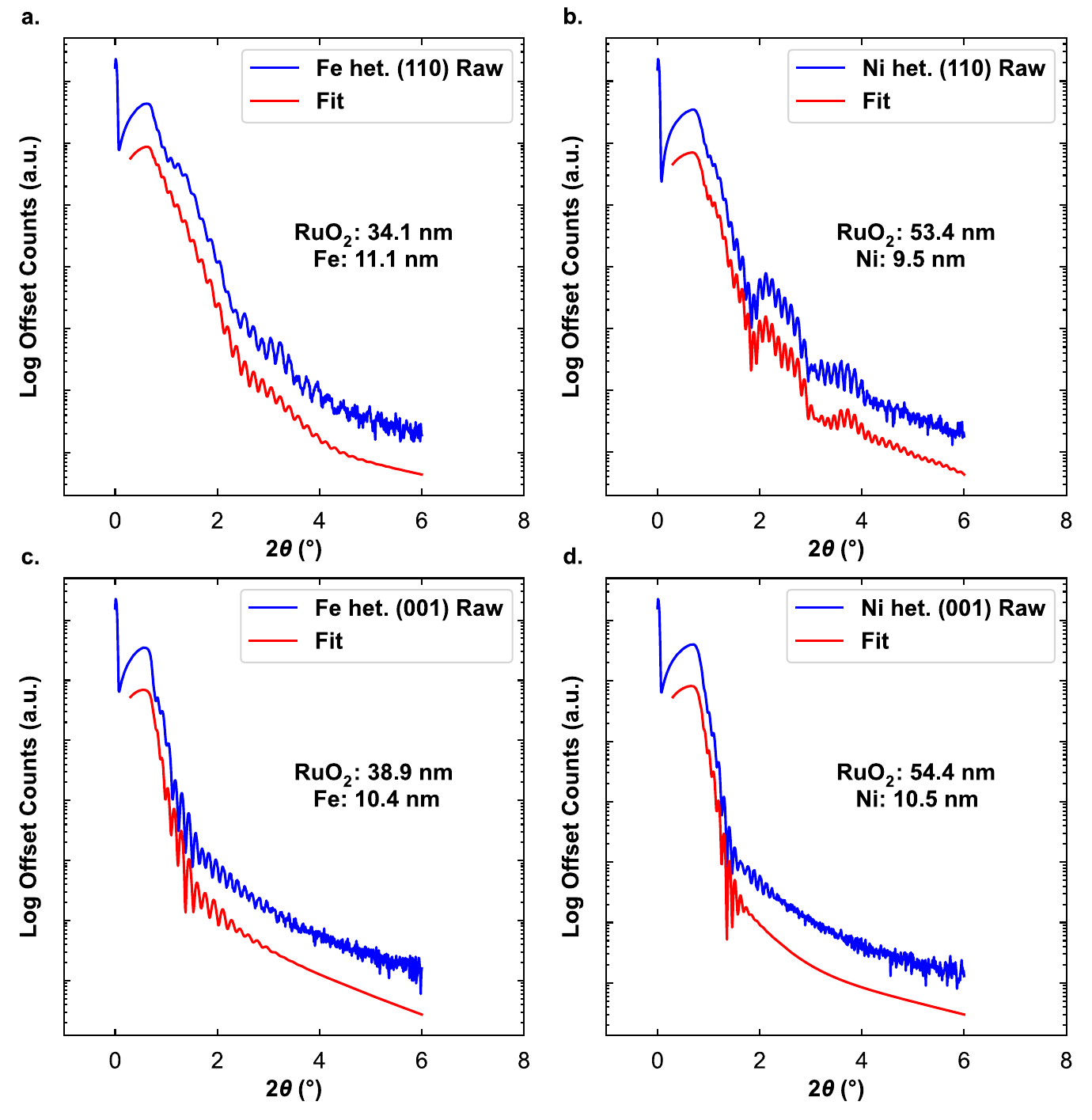}
 \caption{XRR measurements (blue) and offset fitting profiles (red) of heterostructures comprising (a) (110) TiO$_2$/RuO$_2$/Fe/Ru, (b) (110) TiO$_2$/RuO$_2$/Ni/Ru, (c) (001) TiO$_2$/RuO$_2$/Fe/Ru, and (d) (001) TiO$_2$/RuO$_2$/Ni/Ru. In each panel, the fit thicknesses of the RuO$_2$ and FM layers are labeled.}
 \end{figure}

 \begin{figure}[!ht]
 \centering
  \includegraphics[width=0.4\textwidth]{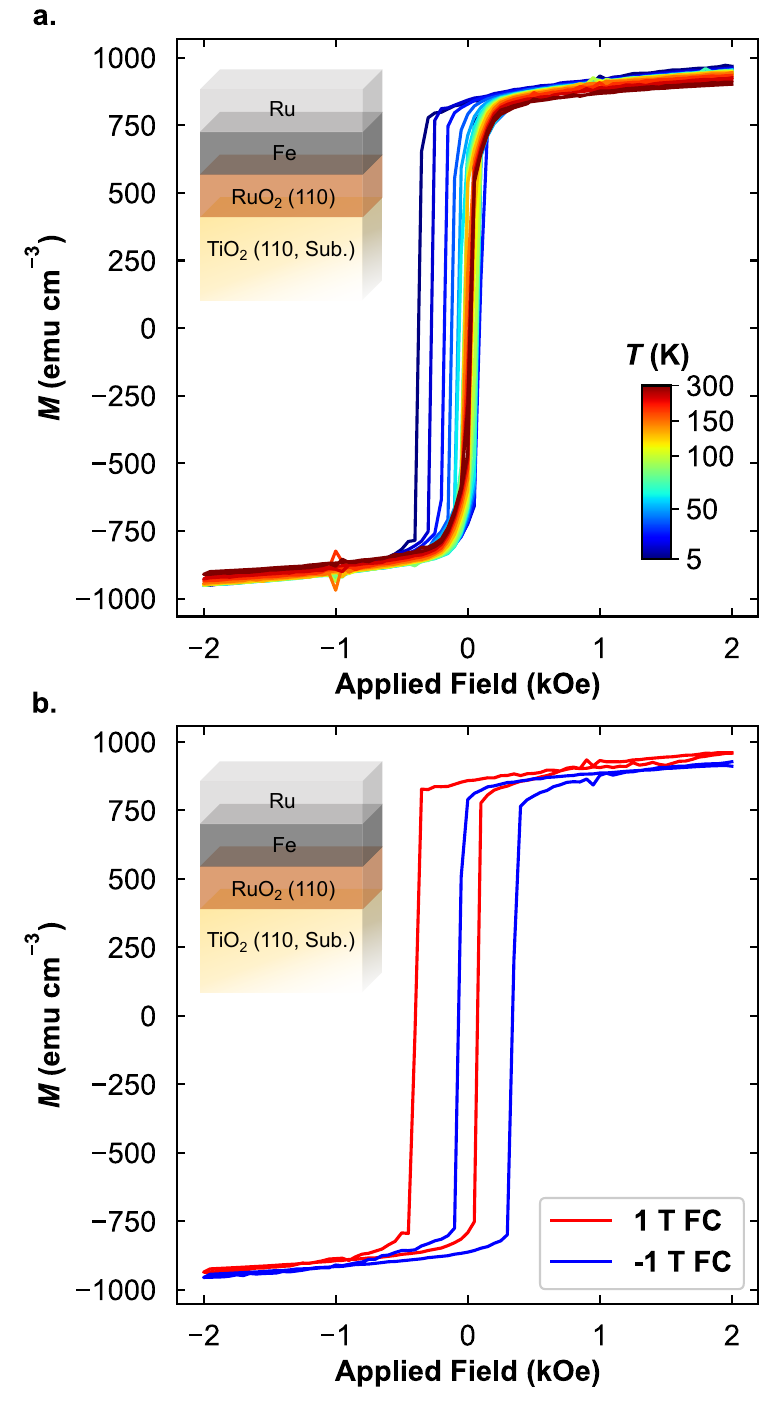}
 \caption{(a) Temperature-dependent in-plane hysteresis loops measured on a (110) TiO$_2$/(110) RuO$_2$/Fe/Ru heterostructure on heating following field cooling from 400~K in a 1~T applied field. The sample is rotated 90$\degree$ in-plane compared to the measurements shown in Figure~2(a). Warmer colors correspond to higher $\it{T}$, whereas cooler colors correspond to lower, as detailed in the inset color bar. (b) Hysteresis loop measurements made at 5~K following cooling from 400~K in 1~T (red) and -1~T (blue) in-plane magnetic fields. The 1~T measurement is also plotted in Figure~2(a). Inset within each panel is an annotated diagram of the measured film stack including TiO$_2$/RuO$_2$ orientation.}
 \end{figure}

\clearpage

A comparison between wavelength-variable room-temperature neutron diffraction measurements of the allowed nuclear ($\bar{3}$$\bar{3}$0) and forbidden (0$\bar{1}$0) peaks from the stacked TiO$_2$ substrates is shown in Figure~S4(a,b). The intensity of the allowed nuclear ($\bar{3}$$\bar{3}$0) peak, shown in Figure~S4(a), displays the expected uniform profile as wavelength is varied between 1.65 and 1.80~\AA. Oppositely, an identical measurement of the forbidden (0$\bar{1}$0) peak, shown in Figure~S4(b), boasts an intensity that is sensitive to wavelength, which is varied between 0.75 and 1.75~\AA, and consequently nonuniform. This observed nonuniformity with wavelength would be expected if the intensity was the result of multiple scattering events, as the initial bragg condition for constructive interference and diffraction cannot be predictably maintained through angular modulation as wavelength is varied. Therefore, this comparison confirms that the forbidden ($\it{h}$00) and (0$\it{k}$0) ($\it{h}$ or $\it{k}$ = odd) peaks present in the white-beam pattern shown in Figure~3(a) arise due to multiple scattering events.

 \begin{figure}[!ht]
 \centering
  \includegraphics[width=0.4\textwidth]{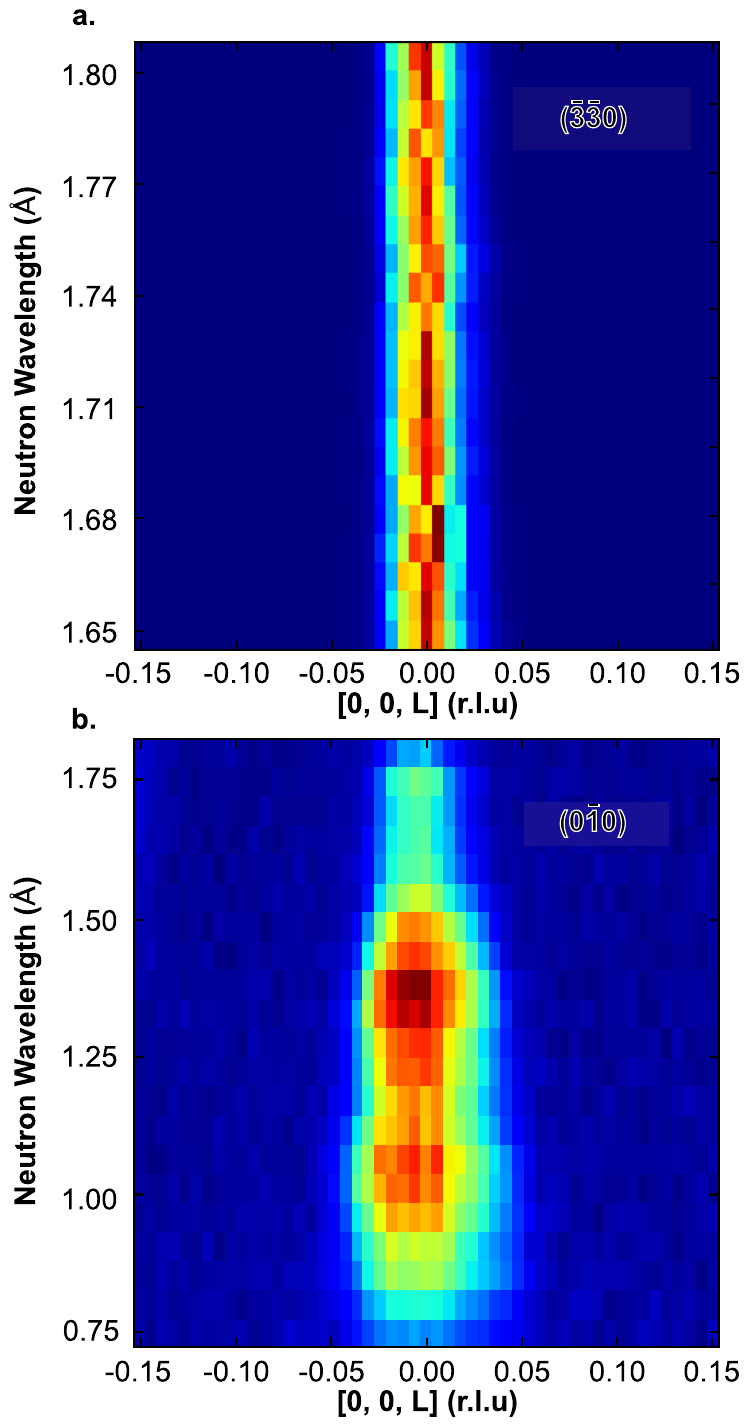}
 \caption{Wavelength-variable room temperature neutron diffraction measurements of (a) the allowed nuclear ($\bar{3}$$\bar{3}$0) and (b) the forbidden (0$\bar{1}$0) peaks from TiO$_2$. Warmer colors correspond to intensity, where each plot is labeled with the corresponding reflection in the upper right.}
 \end{figure}

 \begin{figure}[!ht]
 \centering
  \includegraphics[width=0.4\textwidth]{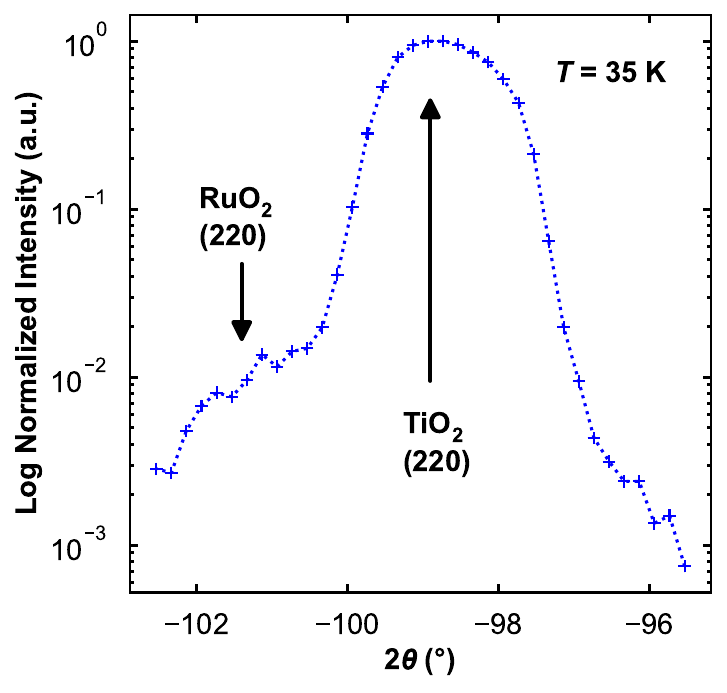}
 \caption{2$\theta$-$\omega$ unpolarized neutron diffraction measurement of the (220) region of TiO$_2$/RuO$_2$ at 35 K showing the proximity of the two diffraction peaks and their relative intensity. The index of each peak is annotated using vertical arrows.}
 \end{figure}

 \begin{figure}[!ht]
 \centering
  \includegraphics[width=0.4\textwidth]{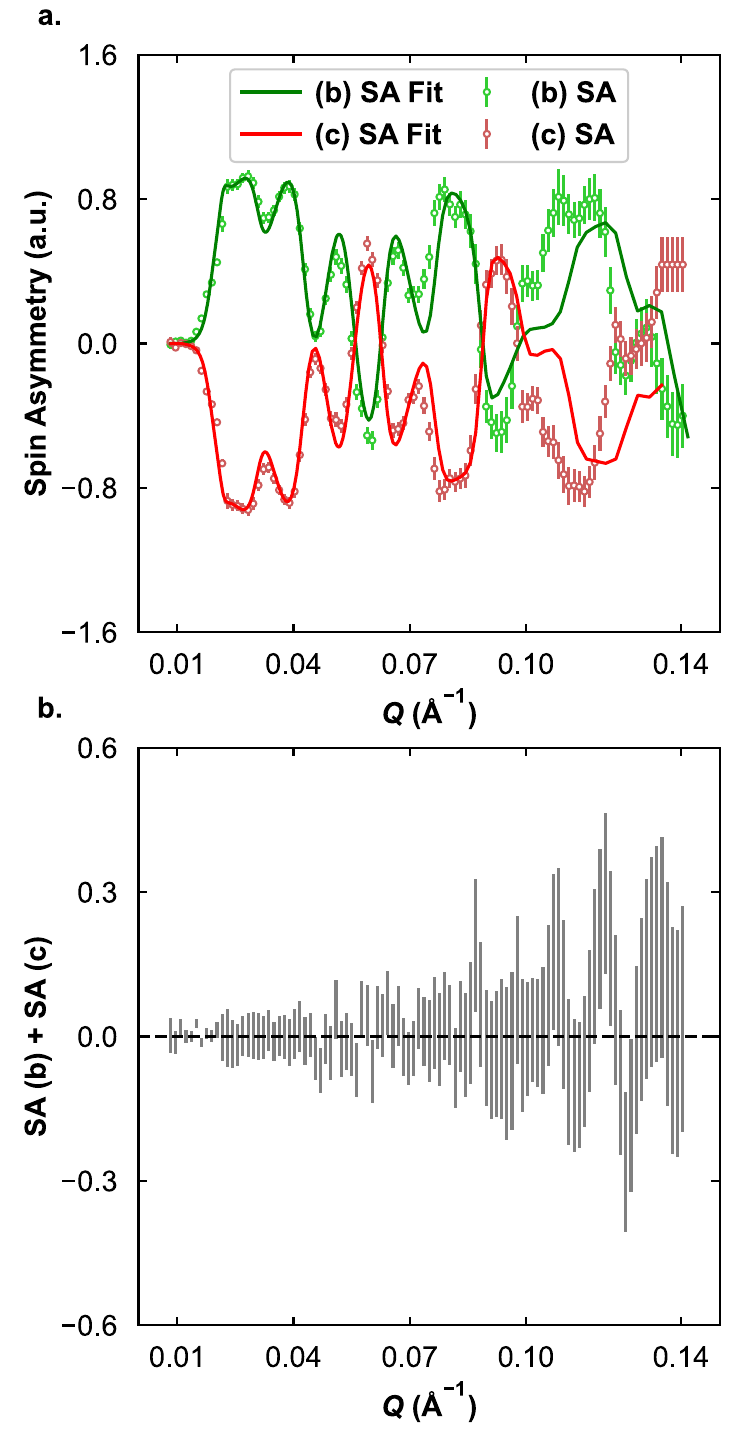}
 \caption{(a) Spin asymmetry data (open points) and fits (lines) from PNR fitting of measurements made on the (110) TiO$_2$/RuO$_2$/Fe/Ru heterostructure  at 50~Oe and 5~K  (green, labeled (b) corresponding to panel (b) in Figure~4) following field cooling from 400~K in a 1~T field and at 50~Oe and 5~K after switching with a -1~T field (red, labeled (c) corresponding to panel (c) in Figure~4) without heating above 5~K. (b) Difference between measured spin asymmetries. In this plot, a black dashed horizontal line is plotted through zero.}
 \end{figure}

 \begin{figure}[!ht]
 \centering
  \includegraphics[width=0.4\textwidth]{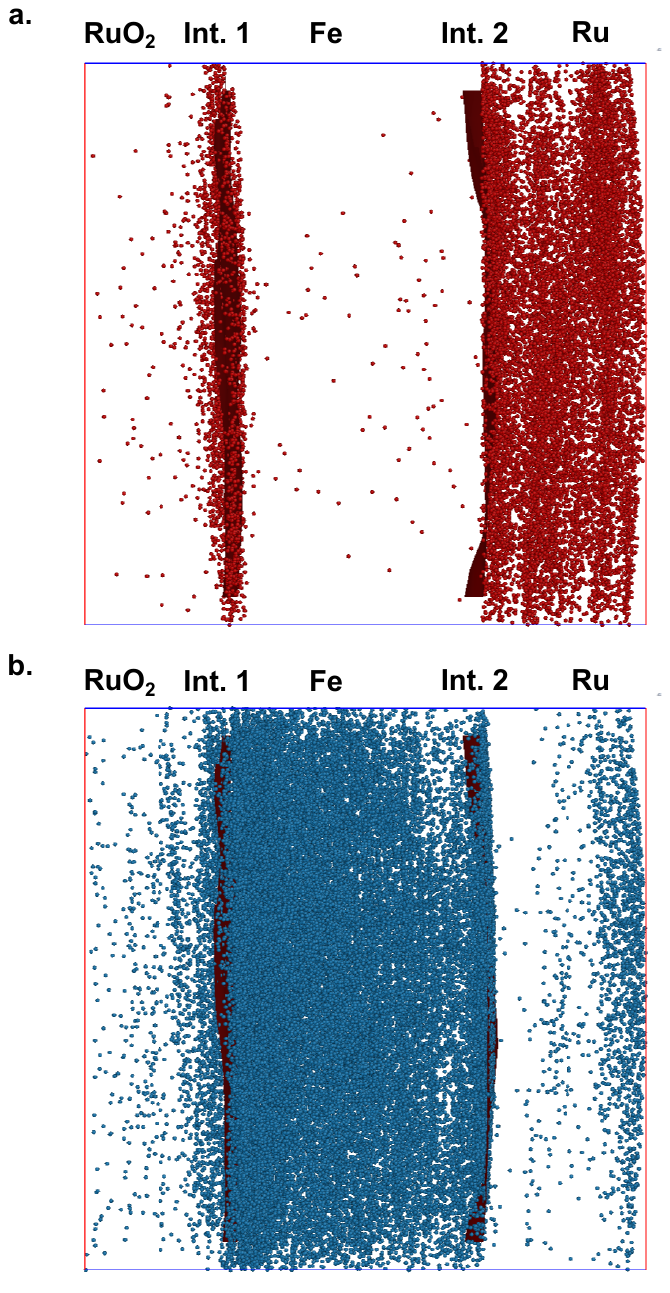}
 \caption{Atom maps for elemental (a) Ru and (b) Fe species within 20~nm-diameter cylindrical ROI collected from tomography milling of (110) TiO$_2$/RuO$_2$/Fe/Ru heterostructure.}
 \end{figure}
\newpage
 \begin{figure}[!ht]
 \centering
  \includegraphics[width=0.4\textwidth]{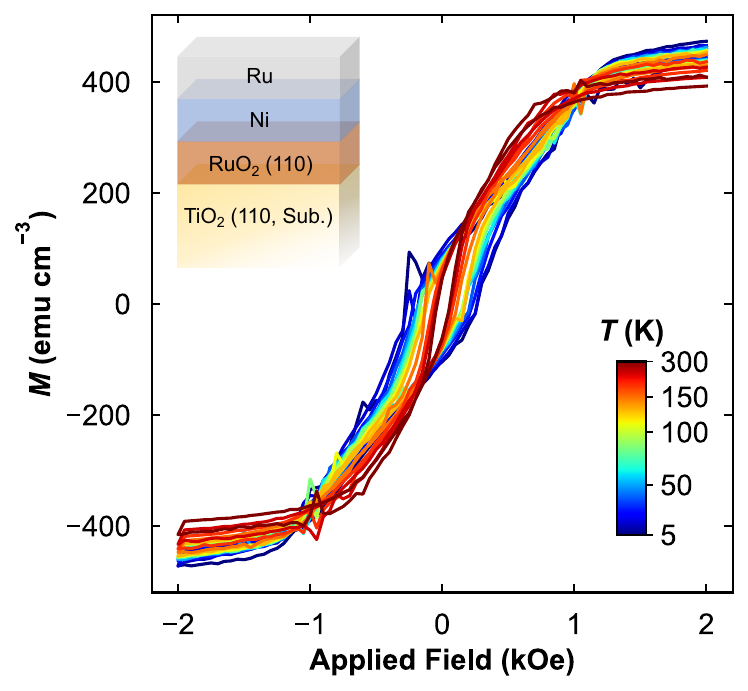}
 \caption{SQUID magnetic hysteresis loops measured on a TiO$_2$ (110)/RuO$_2$/Ni/Ru heterostructure between 5~K and 300~K following the field cooling procedure described in the main text, identical to the measurements plotted in Figure~2(a,b) and Figure~4(a,b). The temperature at which the measurement was made corresponds to loop color, as indicated by the color bar inset in the lower right, and inset in the upper left of the panel is an annotated diagram of the measured film stack including TiO$_2$/RuO$_2$ orientation.}
 \end{figure}

\end{document}